\documentclass[trackchanges,twocolumn]{aastex701}
\shorttitle{Evolution of deformed NSs}
\shortauthors{Biryukov et al.}

\usepackage{amsmath}

\newcommand{\diff}{\ensuremath{\rm d}}

\begin{document}

\title{Rotational evolution of deformed magnetized neutron stars:\\
implications for obliquity distribution and braking indices statistics}

\author[orcid=0000-0001-7210-6988,sname='Biryukov']{Anton Biryukov}
\affiliation{The Raymond and Beverly Sackler School of Physics and Astronomy, Tel Aviv University, Tel Aviv, 6997801, Israel}
\affiliation{Sternberg Astronomical Institute, Lomonosov Moscow State University, 13 Universitetsky pr., Moscow, 119234, Russia}
\email[show]{ant.biryukov@gmail.com}  

\author[sname='Abolmasov', orcid=0000-0002-6429-0436]{Pavel Abolmasov} 
\affiliation{The Raymond and Beverly Sackler School of Physics and Astronomy, Tel Aviv University, Tel Aviv, 6997801, Israel}
\email{}

\author[sname='Levinson', orcid=0000-0001-7572-4060]{Amir Levinson}
\affiliation{The Raymond and Beverly Sackler School of Physics and Astronomy, Tel Aviv University, Tel Aviv, 6997801, Israel}
\email{}


\begin{abstract}
The rotational evolution of a strongly magnetized neutron star (NS), accreting or isolated, is driven by external torques of different nature. 
In addition to the torques, even the tiniest deformations of the NS crust can affect its rotation through asymmetries in its inertia tensor. 
Several factors may be responsible for the deformations, including strong magnetic fields, internal stresses, or 
local heating. The main effect produced by the deformations is the so-called free precession: the motion of the rotational axis 
with respect to the crust. We consider the evolution of a triaxially deformed isolated NS with a strong dipolar magnetic field for a 
broad range of parameters, taking into account the magnetic field decay. We show that the combination of pulsar torques and free precession results in a considerable broadening of the distribution of magnetic obliquity angles (the angle between the magnetic and rotational axes) and creates a population of objects where the rotational axis does not align with the magnetic axis at all but enters a limit-cycle regime. The combination of free precession and magnetic torques can also explain the observed distribution in pulsar braking indices by creating a periodic oscillation in the magnetic obliquity.

%
\end{abstract}

\keywords{\uat{High Energy astrophysics}{739} --- \uat{Magnetic fields}{994} -- \uat{Neutron stars}{1108} --- \uat{Pulsars}{1306}}


\section{Introduction}
Rotation-powered pulsars is a numerous class of isolated neutron stars (NS). They are believed to have a dipolar magnetosphere, whose axis is misaligned with the rotation axis by an angle $\chi$, known as the {\it magnetic angle}.  Its time change, along with the spin frequency $\Omega$, its time derivative $\dot\Omega$, and other higher-order timing parameters, provides a description of the star's rotational evolution.

Radiopulsars are subject to magnetospheric torques and exhibit a gradual spin down over time scales of
$\tau_\mathrm{psr} \sim |\Omega/\dot\Omega| \sim$ millions of years, as expected by theory (\citealt{spitkovsky06}, see also \citet{2024Galax..12....7A} and \citet{2012puas.book.....L} for a review). At the same time, when treated as rigid spherical bodies, misaligned rotators are expected to undergo progressive alignment of the spin and magnetic axes \citep{phil14}, and to precess around the magnetic axis with a period of $\sim(\Omega R_\mathrm{NS}/c)\tau_\mathrm{psr}$, where $R_\mathrm{NS}$ is the neutron star radius.
This period is $10^3-10^4$ times shorter than $\tau_\mathrm{psr}$ for a typical pulsar, and this type of precession is usually referred to as ``radiative precession'' in the pulsar literature \citep{melatos2000, bz_anomal2014}. If, in addition, the inertia tensor of the neutron star is nonisotropic, then other precessional components are at play.

Several mechanisms that might cause biaxial and triaxial deformations of the NS have been discussed so far \citep[e.g.][]{ogunn69, goldreich70, melatos2000, jones_anderssson_2001, arz15}. 
These include: (i) uncompensated elastic deformation that ensues the crystallization of the star \citep{goldreich70, melatos2000}; 
(ii) uneven heating of the neutron star crust due to magnetic field decay or accretion in a binary system  \citep{Bildsten98, usho2000, pons07, pons09}; 
and (iii) distortion of the NS crust by the tension of a sufficiently strong internal magnetic 
field \citep[e.g.][]{haskell08}. 
The magnitudes of the deformation predicted by these models span a wide range, 
from $\varepsilon\sim 10^{-14}$ to $\varepsilon\sim 10^{-5}$, where $\varepsilon = \Delta I/I$, $I$ is the net moment of inertia of the star, and $ \Delta I$ is its variation between different directions. The specific value depends on the details of the NS interior physics.

If $\varepsilon$ is large enough, the star will tend to precess freely with a period $\sim 2\pi/\varepsilon\Omega \ll \tau_\mathrm{psr}$. 
This can manifest itself observationally as periodic changes in the pulsar timing parameters \citep{jones_anderssson_2001}. At least for two isolated pulsars, free precession was proposed as an explanation for a periodicity known from the observations: B1828-11 \citep{akgun06} (see, however, \citealt{stairs19}); and B1642-03 \citep{Shabanova01}. 
The decadal 
increase in the separation between the main and secondary pulses of the Crab pulsar (for $0.62^\circ$ per century, \citealt{lyne13}), which seems to imply a rapid increase of $\chi$, can also be naturally explained by precession \citep{arz15}.  
But perhaps the most compelling evidence for a precession of an NS induced by a strong deformation comes from the recent X-ray polarization measurements of Her X-1 made by the Imaging X-ray Polarimetry Explorer (IXPE, \citealt{2021AJ....162..208S}) probing the spin geometry of the neutron star \citep{Heyl24}. The 35-day variability in the X-ray pulse frequency of Her X-1 also supports the precession \citep{Kolesnikov22}.

In addition, some gamma-ray and fast radio bursts exhibit rapid quasi-periodic oscillations in their light curves, which, although 
debatable, have been attributed to precessional motions of a central magnetar \citep{Levin20,wasserman2022,zhang24}.

Whether NSs can be modeled as rigid bodies is questionable. In reality, the inner crust is coupled to neutron superfluid vortices that are pinned to nuclei in the crustal lattice.
In the case of absolute pinning, free precession is dominated by the superfluid component \citep{Shaham77}. However, in the presence of an external torque, the superfluid can follow the crust precession through thermal vortex creep (\citealt{Alpar87}; cf., \citealt{sedarkin1999}). 
Some evidence of this exists in Her X-1 \citep{Heyl24}. 
Our analysis ignores any coupling to the core and to the superfluid within the crust, and we only briefly touch upon it in Sec. \ref{sec:discuss:super}.

Although only sufficiently strong deformations, $\varepsilon \sim 10^{-8}-10^{-5}$, can cause precessional motions with periods 
of the order of months to decades, as observed in the objects mentioned above, even much smaller deformations, $\varepsilon \lesssim 10^{-10}$, 
can substantially affect the long-term evolution of radio pulsars. 
If common across the pulsar population, they should also affect their statistical properties. 
This motivates a comprehensive evolutionary analysis of a population of deformed NSs.  
This is the primary goal of this paper.
 
Previous studies have focused on spherical stars driven solely by electromagnetic torques or on deformed stars with a negligible contribution of radiative precession \citep[e.g][]{goldreich70, jones_anderssson_2001}. 
Other studies included both, but focused on very specific cases \citep{melatos2000, zanazzi15, arz15}. 
Most attention was paid to very specific combinations of parameters such as biaxial stars, aligned or orthogonal magnetic axis orientation. 
Radiative precession was often ignored or treated in a simplified way. 
No systematic analysis of the parameter space was performed. Magnetic field decay has never been included in the study.

In this paper, we conduct a systematic study of the rotational evolution of isolated NSs for a broad range of parameters, including the possibility of magnetic field decay, and discuss the observational implications. We identify several evolutionary pathways and demonstrate that the long-term evolution of radiopulsars can naturally account for the anomalous braking indices and their correlation with the pulsar age, as observed, as well as a quasi-isotropic distribution of magnetic angles.

The paper is organized as follows. In Section~\ref{sec:model}, we derive and analyze equations that describe the evolution of a deformed precessing neutron star. 
In Section~\ref{sec:results}, the long-term evolution of such a precessing star is considered for a wide range of NS parameters, including the effect of magnetic field decay. Section~\ref{sec:obs} discusses the observational implications. A brief account of the role of neutron superfluid in the inner crust, which is ignored in our analysis, is given in Section~\ref{sec:discuss:super}. We summarize in Section~\ref{sec:summary}.

\section{Physics of neutron star precession}
\label{sec:model}

\subsection{Euler's equations}
\label{sec:model:equations}

\begin{figure}[t]
    \centering
    \includegraphics[width=\columnwidth]{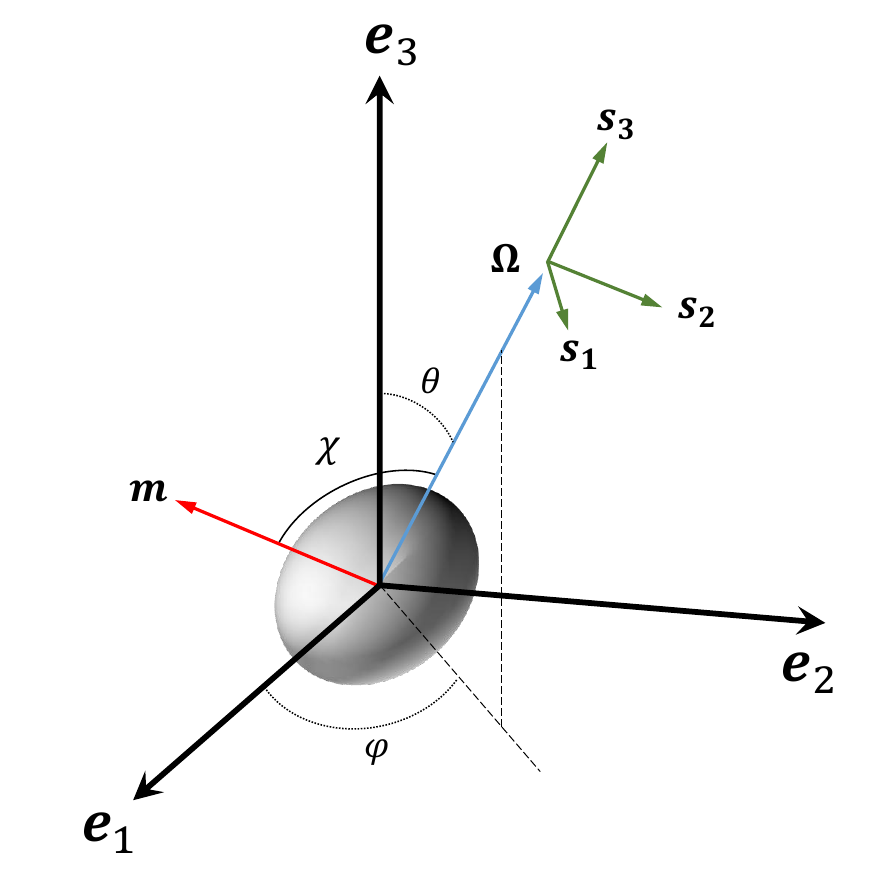}
    \caption{Principal axes of a deformed NS and associated coordinate systems. The angles $\theta$ and $\varphi$ are the polar and azimuthal angles, respectively, measured in the principal frame. The magnetic axis $\pmb m$ is represented by the red arrow. The auxiliary basis $\pmb{s}_{1-3}$ constructed on vectors $\pmb \Omega$ and $\pmb m$ is shown by green arrows.} 
    \label{fig:axes}
\end{figure}

Let us treat the neutron star as a non-spherical rigid body\footnote{The presence of neutron superfluid in the NS interior is ignored in the following analysis; its effect on the rotational evolution will be briefly discussed in Section~\ref{sec:discuss:super}.}, and let $\pmb{e}_k$ ($k = 1..3$) be the unit vectors setting the star's principal axes of inertia. 
In a coordinate system aligned with these axes,  
henceforth referred to as the  {\it principal frame}, the inertia tensor,
\begin{equation}
    I_{ij} = I_0\delta_{ij} + \Delta I_{ij},
\end{equation}
is diagonal. Without loss of generality, the deformation term can be chosen such that $\Delta I_{ij} = \text{diag} (\Delta I_1, 0, \Delta I_3)$, 
with
\begin{equation}
    \Delta I_1 \le 0\mbox{ and }\Delta I_3 \ge 0,
    \label{eq:axes_signs}
\end{equation}
meaning that $\pmb{e}_1$ is the minor principal axis (corresponding to the lowest moment of inertia), $\pmb{e}_3$ is the major principal axis (corresponding to the highest moment of inertia), and $\pmb{e}_2$ corresponds to the intermediate moment of inertia. This configuration is shown in Figure~\ref{fig:axes}. 

The rotational evolution of the star is described by the Euler equations,
\begin{equation}
    \frac{\diff}{\diff t} \pmb{L} = \pmb{L}\times \pmb{\Omega} + \pmb{N},
    \label{eq:Euler_basic}
\end{equation}
where $\pmb{L}$ is the angular momentum, $\pmb{N}$ is the external torque, and
%
%
%
\begin{equation}
    \pmb{\Omega} = 
    \left ( \begin{array}{c}
        \Omega \sin\theta\cos\varphi \\
        \Omega \sin\theta\sin\varphi\\
        \Omega \cos\theta\\
    \end{array} \right ),
    \label{eq:Omega}
\end{equation}
is the angular velocity, where $\theta$ and $\varphi$ are the corresponding polar and azimuthal angles. 

We will use dimensionless parameters $\varepsilon_k = (I_k - I_0)/I_0$, where $k = 1$ or $3$, defined with respect to the intermediate 
moment of inertia, $I_2\equiv I_0$. 
Then the angular momentum takes the form
\begin{equation}
    \pmb L = I_0\pmb \Omega + I_0\pmb \omega,
    \label{eq:angularL}
\end{equation}
where $\pmb \omega = (\varepsilon_1\Omega_1, 0, \varepsilon_3\Omega_3)^T$, with $\varepsilon_1 \le 0$ and $\varepsilon_3 \ge 0$, according to our choice of the principal axes in Eq. (\ref{eq:axes_signs}). 
The vector $\pmb{\omega}$ may be interpreted as a part of the angular momentum related to non-sphericity, per unit moment of inertia $I_0$.
Assuming $I_0 = const$, one can rewrite Euler's equations (\ref{eq:Euler_basic}) as
\begin{equation}
    \pmb{\dot\Omega} + \pmb{\dot\omega} = (\pmb \omega \times \pmb \Omega) + \pmb n.
    \label{eq:Euler}
\end{equation}
Here, the dot denotes the time derivative, and $\pmb n = \pmb N/I_0$.
%
Since the torque acting on an isolated NS is controlled by its magnetic field, it is convenient to treat $\pmb n$ as a sum of three orthogonal terms defined by the relative orientation of the spin and magnetic axes:
\begin{equation}
    \pmb n = \pmb n_\Omega + \pmb n_\chi + \pmb n_\mu = n_\Omega \pmb s_3 + n_\chi \pmb s_2 + n_\mu \pmb s_1.
    \label{eq:torque_decomposed}
\end{equation}
The orthonormal vector basis $\pmb s_i$ ($i = 1,2,3$) is defined in terms of the normalized magnetic moment, $\pmb m = \pmb \mu/\mu$, and the {\it magnetic angle} 
\begin{equation}
    \chi = \cos^{-1} \left( \frac{\pmb m \cdot \pmb{\Omega}}{\Omega} \right)
\end{equation}
as follows:
\begin{equation}
    \pmb s_3 = \pmb \Omega/\Omega,
\end{equation}
\begin{equation}
      \pmb s_2 \sin\chi = (\pmb m \times \pmb s_3) \times \pmb s_3 = \pmb s_3 \cos\chi - \pmb m,
      \label{eq:s2_define}
\end{equation}
and
\begin{equation}
      \pmb s_1 \sin\chi =  (\pmb s_3 \times \pmb m)  = (\pmb s_2 \times \pmb s_3)\sin\chi,
\end{equation}
see also the Figure ~\ref{fig:axes}. The first term on the right-hand side of Eq.~(\ref{eq:torque_decomposed}) is parallel to the spin axis and accounts for the rotational energy losses, i.e., the spin-down itself. The second term is responsible for magnetic angle evolution, while the third term, which is proportional to $n_\mu$, causes rotation of the spin axis around the magnetic axis with the angular velocity of
\begin{equation}
    \pmb{\omega}_\mathrm{rad} = \varepsilon_\mu \Omega\cos\chi \pmb m,
    \label{eq:omega_radiative}
\end{equation}
where
\begin{equation}
    \varepsilon_{\mu} = -\frac{n_\mu}{\Omega^2 \sin\chi\cos\chi}
    \label{eq:epsilon_magnetic_definition}
\end{equation}
is a convenient dimensionless parameter representing the strength of the torque component $\pmb n_\mu$. In the pulsar literature, this type of precession is often referred to as ``radiative'' precession, and the torque $\pmb n_\mu$ as the ``anomalous'' or ``radiation-precession'' torque \citep{goldreich70, good85, melatos2000}. We shall henceforth refer to this type of forced precession as radiative.

With the decomposition (\ref{eq:torque_decomposed}) and definition (\ref{eq:omega_radiative}), one can re-write Euler's equations (\ref{eq:Euler}) in the form
\begin{equation}
    \dot {\pmb \Omega} + \dot{\pmb \omega} = \left (\pmb \Omega_\mathrm{p} \times \pmb \Omega \right ) + \pmb n_{\chi} + \pmb n_\Omega,
    \label{eq:Euler_final}
\end{equation}
where 
\begin{equation}
    \pmb \Omega_\mathrm{p} = \pmb \omega + \pmb \omega_\mathrm{rad}
    \label{eq:Omega_precession}
\end{equation}
is the full precessional angular velocity. The precessional time scale is $T_\mathrm{p} \sim 2\pi/\Omega_\mathrm{p}$, respectively. Projecting Eq.~\eqref{eq:Euler_final} on $\pmb s_3$ and $\pmb s_2$, one can obtain a set of two equations describing, respectively, the spin down and magnetic angle evolution of the neutron star.

\subsection{Pulsar torque}
\label{sec:model:torque}

The components of the torque $\pmb n$ acting on an isolated active radiopulsar have been obtained numerically and analytically. Generally, this torque is proportional to
\begin{equation}
    n_\mathrm{psr} = -\frac{\mu^2}{I_0 R_\mathrm{LC}^3},
\end{equation}
where $R_\mathrm{LC} = c/\Omega$ is the radius of the light cylinder. Recent MHD and PIC studies of pulsar magnetospheres \citep{spitkovsky06, tche13,phil14}, yield the following torque components:
\begin{equation}
    n_\Omega = n_{\mathrm{psr}} (k_0 + k_1\sin^2\chi),
    \label{eq:N_Omega_iso}
\end{equation}
and
\begin{equation}
    n_\chi = k_2 n_{\mathrm{psr}} \sin\chi\cos\chi.
    \label{eq:N_chi_iso}
\end{equation}
The dimensionless coefficients $k_i$ in these expressions were also estimated numerically and found to be of order unity, with a small spread within the interval 1..2, depending on the particular setup \citep{phil15, Petri20}. Hereafter we adopt $k_0 = k_1 = k_2 = 1$.

The precessional component of the torque, $n_\mu$, has also been found to be non-zero for an isolated neutron star. It has been derived earlier \citep[e.g.][]{dg70, good85, melatos2000, bz_anomal2014} and may be interpreted, at least partially, as a manifestation of the inertia of the stellar magnetic field. 
The most complete consideration of it was provided in the analytical work by \citet{bz_anomal2014}, where the authors considered a spherical NS in a vacuum and found that
\begin{equation}
    n_\mu = - \xi n_{\mathrm{psr}} \frac{R_{\mathrm{LC}}}{R_{\mathrm{NS}}} \sin\chi\cos\chi.
    \label{eq:N_mu_iso}
\end{equation}
The factor
\begin{equation}
    \xi = \frac{8}{15} - \frac{1}{5} \frac{R_{\mathrm{NS}}}{R_{\rm in}}
\end{equation}
was calculated under the assumption that the field is uniform within some radius $R_{\rm in} \le R_{\mathrm{NS}}$ and dipolar outside it. Here, we adopt $R_{\rm in} = R_{\mathrm{NS}}$, for which $\xi = 1/3$, resulting in
\begin{equation}
    \varepsilon_\mu = - \frac{\mu^2}{3 I_0 R_{\mathrm{NS}} c^2},
\end{equation}
independent of $\Omega$ and $\chi$. For a star with an equatorial magnetic field $B_{12} = B / 10^{12}\,$G, radius $R_\mathrm{NS,12} = R_\mathrm{NS}/12\,$km, 
and moment of inertia $I_{0,45} = I_0/10^{45}\,{\rm g\,cm^2}$,
\begin{equation}
    \varepsilon_\mu \approx -9.2\times 10^{-13} B^2_{12} I_{0,45}^{-1} R_\mathrm{NS,12}^{5}.
    \label{eq:epsmu_norm}
\end{equation}
This parameter corresponds to the effective deformation an NS should have in order to achieve the same precession rate without electromagnetic torques. In principle, it can be estimated from the NS timing parameters, simply assuming $\dot\Omega = n_\mathrm{psr}$. In this case,
\begin{equation}
    \hat{\varepsilon}_\mu = -2.1\times 10^{-13} R_\mathrm{NS,12}^{-1} P_\mathrm{s} \dot P_{-15},
    \label{eq:hat_epsilon_mu}
\end{equation}
where the spin period $P_\mathrm{s} = P/(1\mbox{ s})$ and the period derivative $\dot P_{-15}$ is normalized to $10^{-15}$ s s$^{-1}$.
Comparison of Eqs~(\ref{eq:N_Omega_iso}) and (\ref{eq:N_chi_iso}) to Eq.~(\ref{eq:N_mu_iso}) shows that the torque component $n_\mu$ is typically stronger than the other two by a factor
\begin{equation}
    \frac{R_{\mathrm{LC}}}{R_{\mathrm{NS}}} \approx 3.8\times 10^3 P_\mathrm{s},
\end{equation}
which is larger than unity even for young pulsars with $P < 0.1$ s. This means that the radiative precession
timescale,
\begin{figure}[t]
    \centering
    \includegraphics[width=\columnwidth]{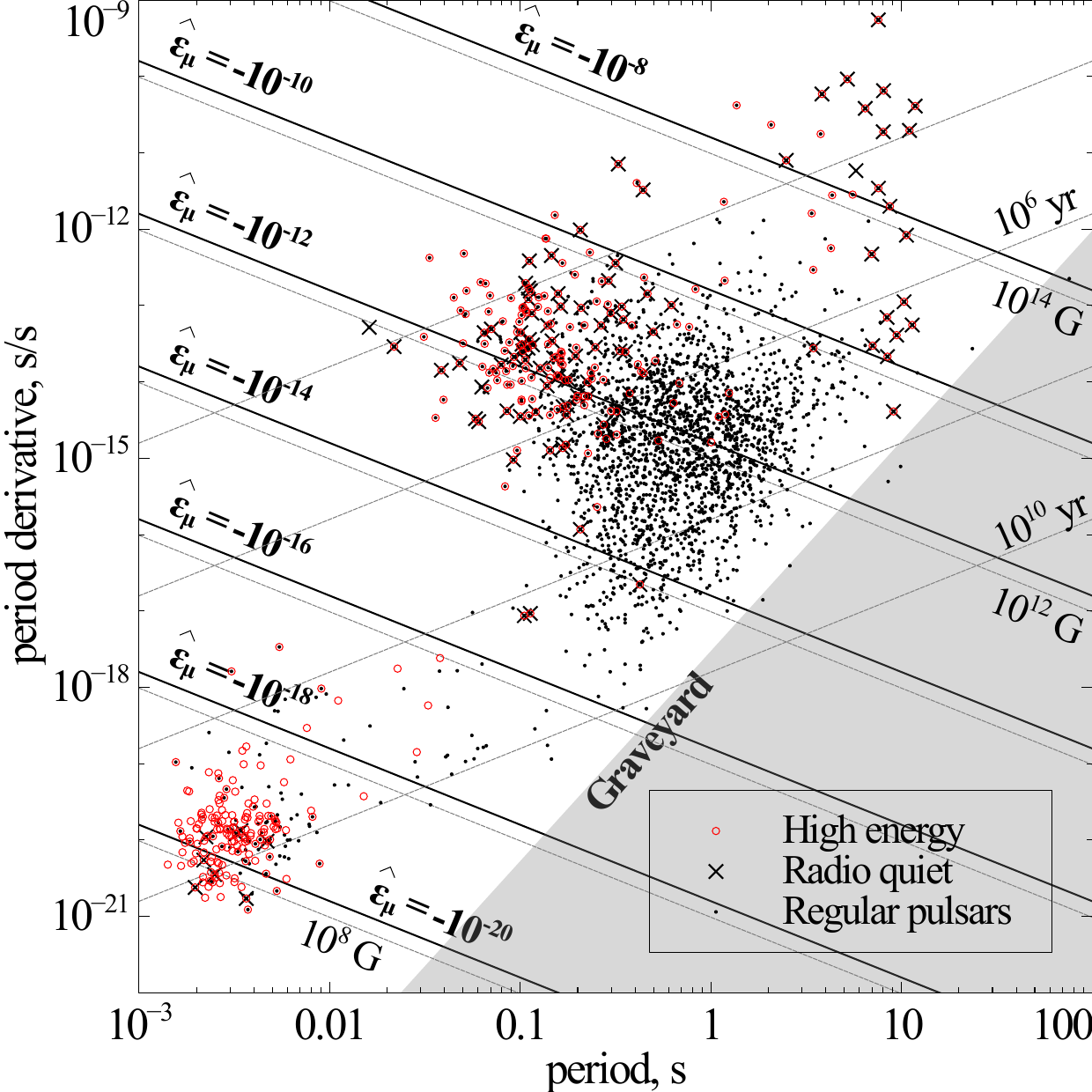}
\\    \caption{The classical period -- period derivative diagram for known isolated radiopulsars. Levels of constant effective radiative deformation $\hat\varepsilon_\mu$ are shown by solid black lines. Gray dashed lines represent constant magnetic fields, obtained from the relation $B_\mathrm{md} = 3.2\times 10^{19} \sqrt{P \dot P}$ G, and constant age, $\tau_\mathrm{ch} = P/2\dot P$. Assuming that all the NSs have comparable deformations $\varepsilon_\mathrm{d}$, one concludes that precession of highly magnetized stars ($B_\mathrm{md} \sim 10^{13}..10^{14}$ G) is dominated by the radiative torque, even if $\varepsilon_\mathrm{d} \sim 10^{-10}..10^{-9}$. On the other hand, even a small deformation $\varepsilon_\mathrm{d} \sim 10^{-17}..10^{-16}$ is sufficient to make free precession dominant in a millisecond pulsar rotation.} 
    \label{fig:ppdot}
\end{figure}
\begin{equation}
    \tau_\mathrm{rad} = - \frac{2\pi}{\varepsilon_\mu \Omega} \approx (3.4\times 10^4\mbox{ yr}) P_\mathrm{s} B_{12}^{-2} I_{0,45} R_\mathrm{NS,12}^{-5},
    \label{eq:tau_rad}
\end{equation}
is {\it always} shorter than the pulsar spin-down timescale,
\begin{equation}
    \tau_\mathrm{psr} = - \frac{\Omega}{n_\mathrm{psr}} \approx (7.3\times 10^6\mbox{ yr}) P_\mathrm{s}^2 B_{12}^{-2} I_{0,45} R_\mathrm{NS,12}^{-6}.
    \label{eq:timescale_pulsar}
\end{equation}
\begin{figure*}
    \centering
    \includegraphics[width=\textwidth]{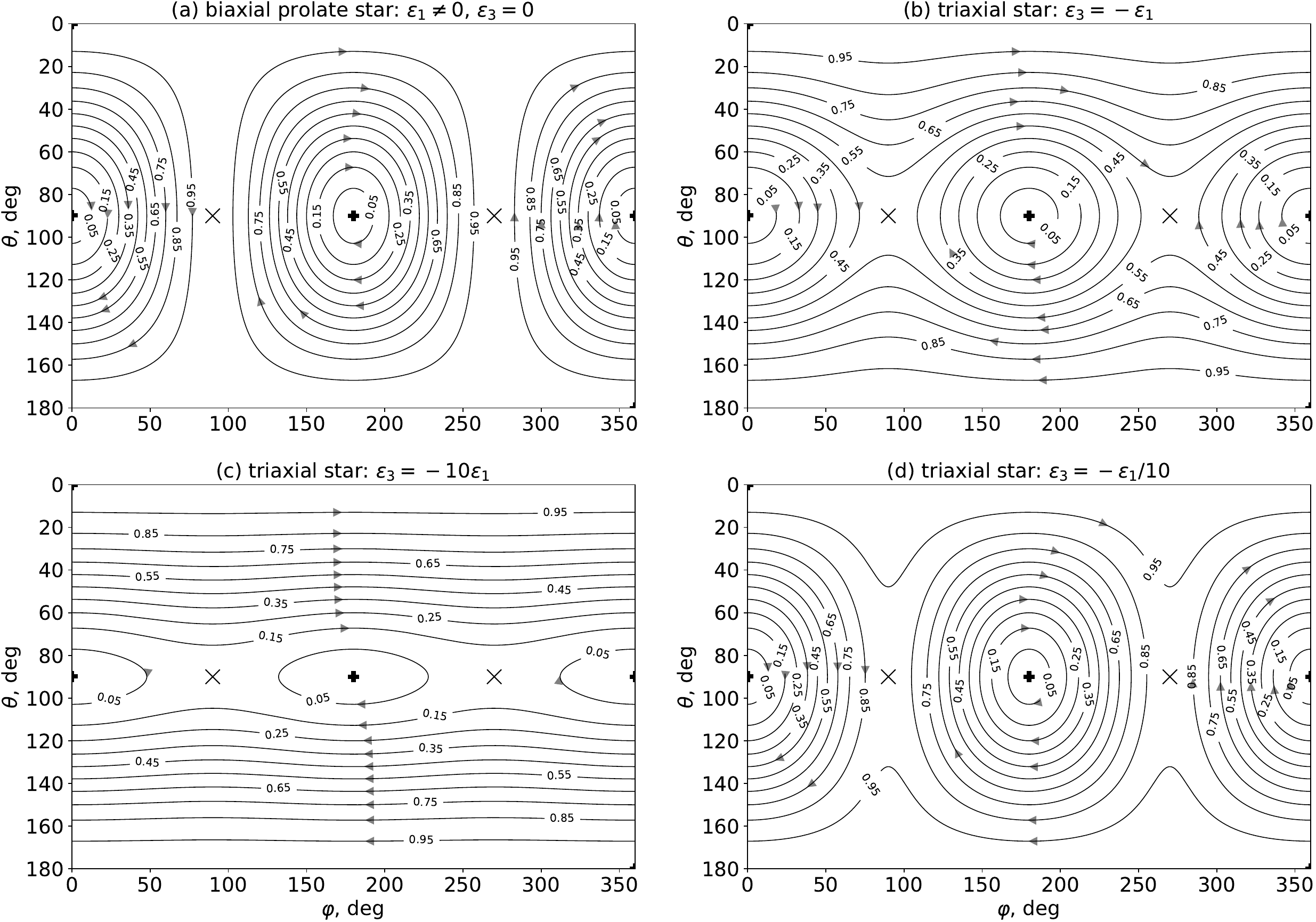}
    \caption{Set of trajectories of the spin axis for a freely precessing star, represented by the contours of $\lambda_\mathrm{0,norm} = const$. The numbers that label the curves are values of $\lambda_{\rm 0,norm}$.  Four different values of the ratio $\varepsilon_{3}/\varepsilon_1$ are shown.  In each case the spin vector $\pmb \Omega$ follows one of the contour lines ({\it the polhodes}) with precession period $\sim 2\pi/\varepsilon_\mathrm{d}\Omega$. The bold plus signs at the centers of the closed sets of contours correspond to minima of $\lambda_0$. The maxima occur at the polar angles $\theta = 0$ and $180^\circ$. The crosses depict the positions of the saddle points (X-points). Case (a) with $\varepsilon_{3} = 0$ represents a prolate biaxial deformation along the $\pmb{e}_1$ axis. In case (b) both deformations have the same amplitude: $\varepsilon_1 = -\varepsilon_3$. The other two cases, $\varepsilon_3/\varepsilon_1 = -10$ and $\varepsilon_3/\varepsilon_1 = -0.1$, correspond to deformations predominantly along $\pmb{e}_3$ and $\pmb{e}_1$, respectively.} 
    \label{fig:lambda_free_precession}
\end{figure*}
Consequently, even a spherical neutron star is always subject to precessional motion, with its spin axis undergoing many revolutions around the precessional axis before the star reaches its death line. If the star has a non-negligible deformation,
\begin{equation}
    \varepsilon_\mathrm{d} \equiv \sqrt{\varepsilon_1^2 + \varepsilon_3^2} ,
\end{equation}
this conclusion remains true, though the precession frequency is different. 
If the deformation is strong enough $\varepsilon_\mathrm{d}\gg |\varepsilon_\mu|$, it dictates the precessional time: 
\begin{equation}
    T_\mathrm{p} (\varepsilon_\mathrm{d} \gg |\varepsilon_\mu|) \approx \frac{2\pi}{\varepsilon_\mathrm{d}\Omega} \simeq (3.2\times 10^3\mbox{ yr}) \left( \frac{\varepsilon_\mathrm{d}}{10^{-11}}\right)^{-1} \,P_\mathrm{s}.
    \label{eq:precess_timescale}
\end{equation}
Radiative precession alone does not affect the magnetic angle evolution and has a weak effect on the observed spin-down rate \citep{jones_anderssson_2001, bbk12}. 
However, when coupled with biaxial or triaxial deformations, it can produce a significant observational impact, as discussed earlier by \citet{melatos2000} and further considered in the following sections.

It is worth noting that because of the wide range of pulsar magnetic fields, $\varepsilon_\mu$ has a large scatter across the pulsar population. In Figure \ref{fig:ppdot}, the period-period derivative diagram is shown for known isolated neutron stars\footnote{The data were taken from the ATNF pulsar database \citep{atnf}, URL:\texttt{https://www.atnf.csiro.au/research/pulsar/psrcat/}, v.2.6.0}.
The lines of constant magnetic field, $B_\mathrm{md} = 3.2\times 10^{19} \sqrt{P \dot P}$\,G, and characteristic age, $\tau_\mathrm{ch} = P/2\dot P$, are overplotted. 
In addition, regions with different $\hat{\varepsilon}_\mu$ are shaded with different colors. If the deformation of the NS exceeds $\varepsilon_\mu$ at a certain point in the diagram, free precession dominates over the radiative one.
Thus, a highly magnetized neutron star would be precessing mostly due to pulsar torques even if the deformation is as large as $\varepsilon_{\rm d} \sim 10^{-10}..10^{-9}$. On the other hand, even a small deformation, $\varepsilon_\mathrm{d} \sim 10^{-17}..10^{-16}$, is sufficient to render free precession dominant in the rotation of a millisecond pulsar. 
The balance between free and radiative precession is capable of making the rotational evolution of these types of pulsars significantly different. 

\begin{figure*}
    \centering
    \includegraphics[width=\textwidth]{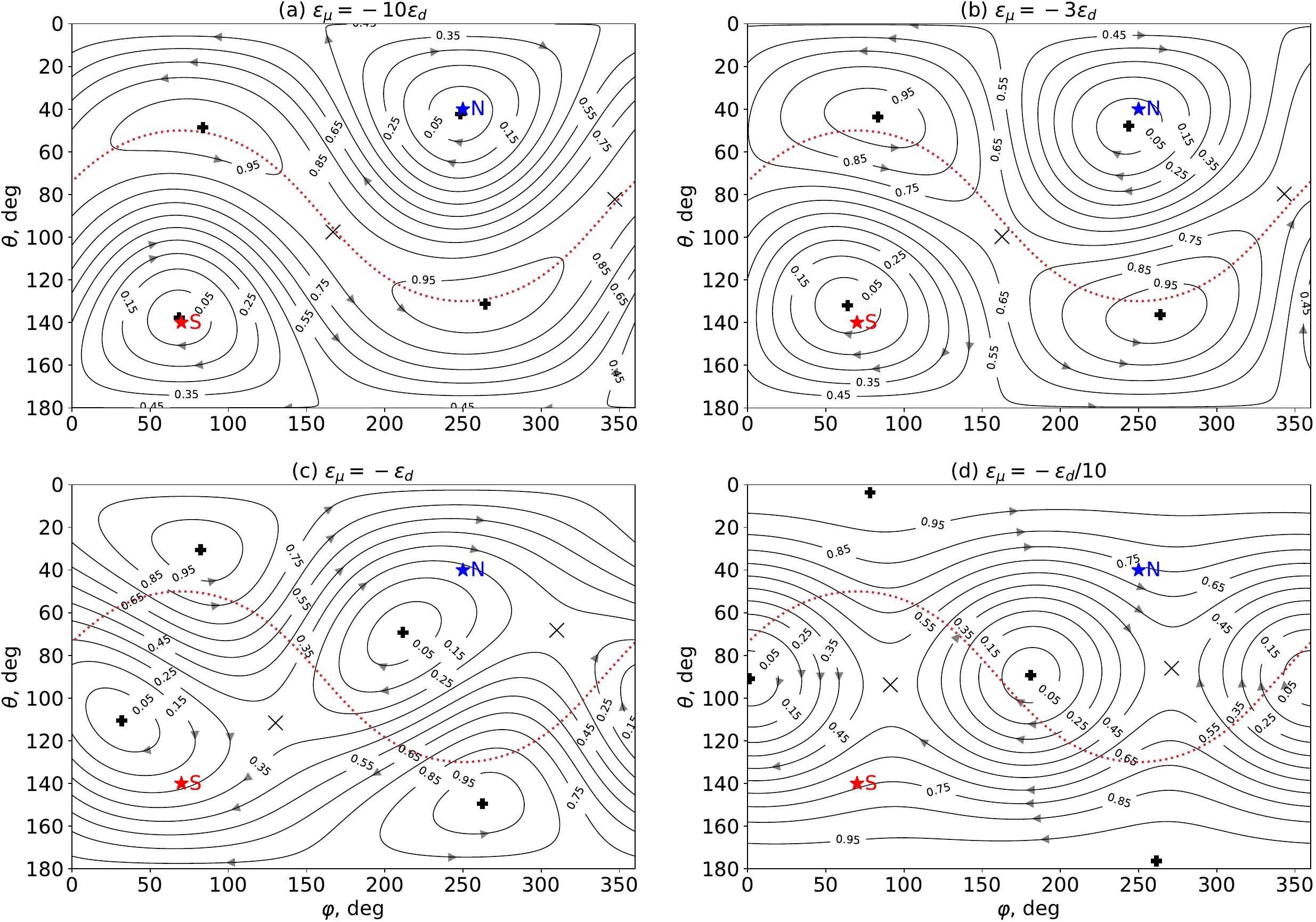}
    \caption{Precession trajectories (contours of $\lambda_\mathrm{norm})$ for a triaxial star with $\varepsilon_1 = -\varepsilon_3$ and nonzero radiative torque $n_\mu$ with ratios $\varepsilon_\mu/\varepsilon_\mathrm{d} = -0.1,-1,-3$ and $-10$.
    In all plots, the north magnetic pole has the coordinates $\varphi_\mathrm{m}=250^\circ$, $\theta_\mathrm{m} = 50^\circ$. For different orientations of the magnetic axis, the configurations are qualitatively the same. The north (N) and south (S) magnetic poles are represented by blue and red stars, respectively. The dotted red line depicts the magnetic equator, while other lines and marks have the same meaning as in Figure~\ref{fig:lambda_free_precession}.} 
    \label{fig:lambda_total_precession}
\end{figure*}

\subsection{Free precession}
For a freely precessing star ($\pmb{n} \equiv 0$), Eq.~(\ref{eq:Euler_final}) reduces to
\begin{equation}
    \dot{\pmb{\Omega}} + \dot{\pmb{\omega}} = (\pmb{\omega} \times \pmb{\Omega}).
\end{equation}
Projecting the resultant equations onto $\pmb{\omega}$ leads to a constant of motion 
\begin{equation}
    \lambda_0 = \frac{\pmb{\Omega}\cdot\pmb{\omega}}{2\Omega^2} = \frac{1}{2}\left(\varepsilon_1 \sin^2\theta \cos^2\varphi + \varepsilon_3 \cos^2\theta\right).
    \label{eq:lambda_free}
\end{equation}
Here we have assumed that $\varepsilon_{1,3} \ll 1$ and both remain constant. These assumptions will be used further in the text.

The geometric meaning of $\lambda_0$ is as follows. The tip of the spin vector $\pmb\Omega$ rotates in the principal frame along trajectories of constant $\lambda_0$ ({\it the polhodes}) with a precessional period $T_\mathrm{p} \sim 2\pi/\varepsilon_\mathrm{d} \Omega$. 
The set of available precession orbits depends on the ratio between $\varepsilon_1$ and $\varepsilon_3$. 
In Figure~\ref{fig:lambda_free_precession}, we show examples of the precessional orbits identified by their normalized $\lambda_0$ parameter values
\begin{equation}
    \lambda_\mathrm{0,norm} = \frac{\lambda_0 - \lambda_\mathrm{0,min}}{\lambda_\mathrm{0,max} - \lambda_\mathrm{0,min}} \in [0,1],
    \label{eq:lambda_norm}
\end{equation}
with $\lambda_\mathrm{0,min}$ and $\lambda_\mathrm{0,max}$ being the minimal and maximal values of $\lambda_0$ for a particular choice of the deformation parameters $\varepsilon_1, \varepsilon_3$. 

It is evident from Eq. (\ref{eq:lambda_free}) that there are four precession poles (extrema of $\lambda_0$): two along the minor principal axis $\pmb{e}_1$, and two along the major principal axis $\pmb{e}_3$. They are depicted by bold plus signs in Figure \ref{fig:lambda_free_precession}. With our convention, $\varepsilon_1 < 0$, the corresponding values of $\lambda_0$ at these locations are $\lambda_{0,\mathrm{min}} = \varepsilon_1/2$ and $\lambda_{0,\mathrm{max}} = \varepsilon_3/2$. The saddle points (X-points), that correspond to $\lambda_0 = 0$, are located on the intermediate axis $\pmb{e}_2$ and are marked by crosses. In the case of an oblate ($\varepsilon_1 = 0$) or prolate ($\varepsilon_3 = 0$) biaxial deformations, $\lambda_{0,\mathrm{min}}$ or $\lambda_{0,\mathrm{max}}$ are equal to zero, respectively.

Generally, the spin axis tends to precess around the axis of minimal or maximal inertia, while the intermediate axis $\pmb{e}_2$ remains unstable. This is a well-known result of classical mechanics.

\subsection{The combined effect of the two precessional motions}

A real neutron star is expected to have a nonzero radiative precessional torque $n_\mu$, which contributes to the full precessional angular velocity (Eq.~\ref{eq:Omega_precession}). In this case, one can introduce a more general version of the scalar $\lambda_0$,
namely\footnote{The scalar $\lambda$ is the generalization of the parameter $\gamma$ used by \citet{melatos2000}, see Eq. (20) of that paper.}.
\begin{equation}
    \lambda = \frac{\pmb \Omega \cdot \pmb \Omega_\mathrm{p}}{2\Omega^2} = \frac{1}{2}\left(\varepsilon_1 \sin^2\theta \cos^2\varphi + \varepsilon_3 \cos^2\theta + \varepsilon_\mu \cos^2\chi\right).
    \label{eq:lambda_define}
\end{equation}
As will be shown below, if $\varepsilon_\mu = const$ and $\pmb{m}$ is fixed in the principal frame, $\lambda$ evolves on time scales $\sim \tau_\mathrm{psr}$ much longer than the precessional period $\sim 2\pi/\Omega_\mathrm{p}$. Therefore, over short enough times, the precession still follows $\lambda$ contours to a good approximation. 

Generally, radiative precession changes the lines of constant $\lambda$ in a way that depends on the ratio between $\varepsilon_\mathrm{d}$ and $\varepsilon_\mu$. In particular, if $|\varepsilon_\mu| \gg \varepsilon_\mathrm{d}$, the NS spin axis will tend to precess mostly around the magnetic axes rather than one of the principal axes. 

Figure~\ref{fig:lambda_total_precession} shows the shapes of the precession orbits for different relative contributions of the free and radiative precession terms. In this plot, the magnetic axis is tilted in the direction $\varphi_\mathrm{m}=250^\circ$, $\theta_\mathrm{m} = 50^\circ$, as measured in the principal frame, but the behavior is quite similar for other choices of initial tilt.
Figure~\ref{fig:lambda_total_precession}(a) depicts contours of constant $\lambda$  for a triaxial star with $\varepsilon_3 = -\varepsilon_1$ and $\varepsilon_\mu/\varepsilon_\mathrm{d} = -10$. 
The values of $\lambda_\mathrm{norm}$ shown there were calculated similarly to $\lambda_\mathrm{0,norm}$. 
The north (N) and south (S) magnetic poles are shown by the blue and red stars, respectively; the dotted red line depicts the magnetic equator, while other lines have the same meaning as in Figure~\ref{fig:lambda_free_precession}. 
Interestingly, even for such high values of $|\varepsilon_\mu|/\varepsilon_\mathrm{d}$, centers of precession orbits are still noticeably shifted from the magnetic poles. 
Moreover, there are two poles of precession around maximal $\lambda$ located near the magnetic equator. In particular, Figure~\ref{fig:lambda_total_precession}(b) shows the case of $\varepsilon_\mu/\varepsilon_\mathrm{d} = -3$, 
Figure~\ref{fig:lambda_total_precession}(c) is for $\varepsilon_\mu/\varepsilon_\mathrm{d} = -1$, while 
Figure~\ref{fig:lambda_total_precession}(d) is for $\varepsilon_\mu/\varepsilon_\mathrm{d} = -0.1$. 
As the effect of radiative precession becomes weaker, the contours become similar to the free-precession setup shown in Figure~\ref{fig:lambda_free_precession}(b).

Even in the presence of effective deformation created by magnetic fields, the $\lambda$ map retains its general triaxial structure with four poles and two X-points.

\section{Long-term evolution of a precessing neutron star}
\label{sec:results}

\subsection{Overview}
\begin{figure*}[t]
    \centering
    \includegraphics[width=\textwidth]{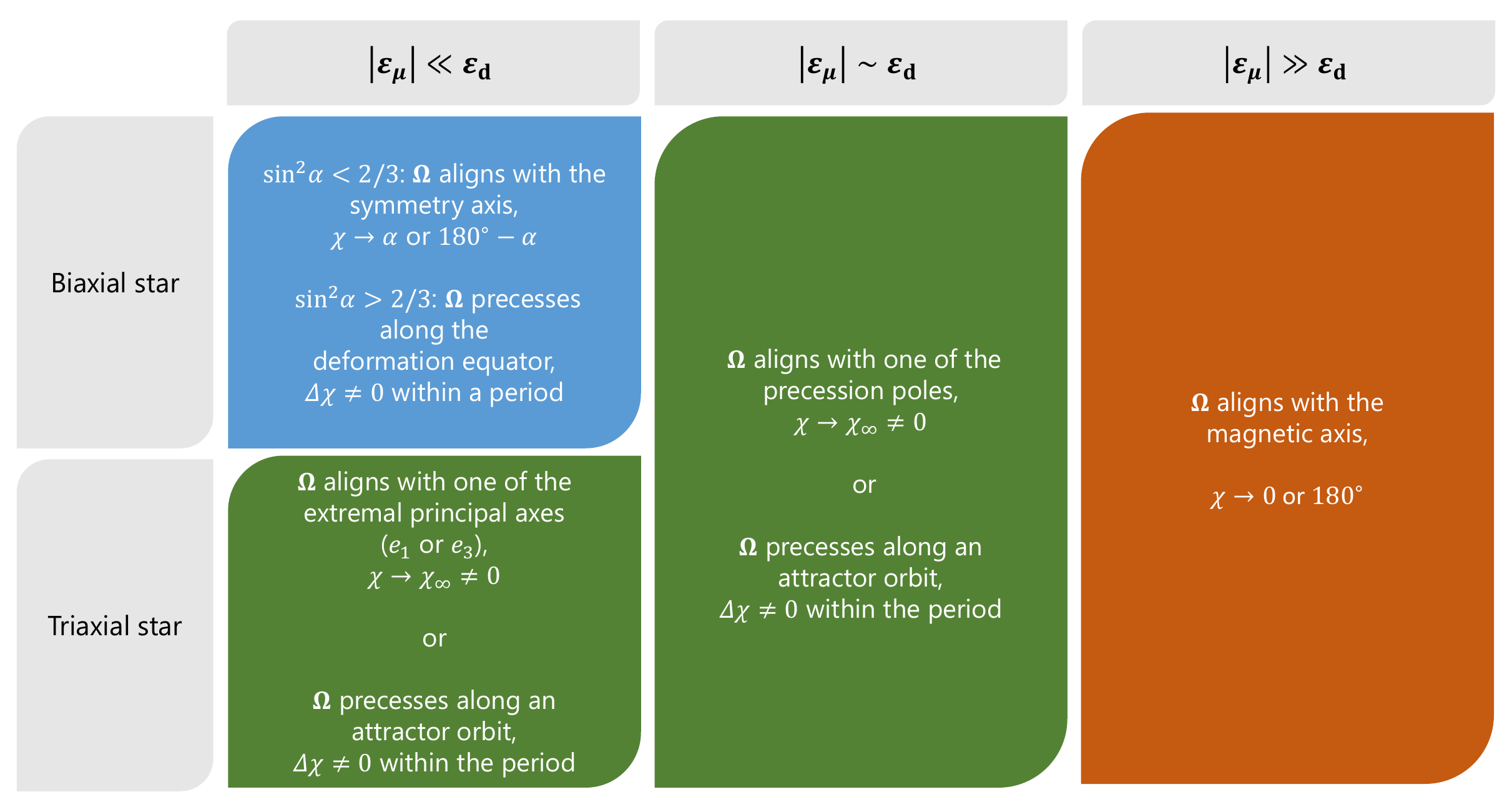}
    \caption{Qualitative summary for asymptotic evolutionary states of a precessing magnetized neutron star. Here, $\pmb\Omega$ denotes the angular velocity vector and $\chi$ the magnetic angle. For a biaxial star, $\alpha$ is the angle between the magnetic and symmetry axes.} 
    \label{fig:evol_cheme}
\end{figure*}
A precessing NS evolves on time scales much longer than its precession period due to the other torque components and secular evolution of the parameters of the star. 
After considerable time, its precession orbits evolve towards either static points or attractor limit circles.
We begin with a qualitative description of the main results.  Readers who are not interested in the gory details can skip the rest of this section. The main observational consequences will be discussed in Sec. \ref{sec:obs}.

The long-term evolution of the precessional motion of a NS depends on the ratio $|\varepsilon_\mu|/\varepsilon_\mathrm{d}$, the orientation of $\pmb{m}$, and the initial orientation of the rotation axis $\pmb{s}_3$ with respect to the principal frame. It takes place on the pulsar spin-down time scale, 
\begin{equation}
    \tau_\mathrm{psr} \gg T_\mathrm{p} \gg P.
\end{equation}
Generally, three regimes can be identified (see Figure~\ref{fig:evol_cheme}). When $|\varepsilon_\mu| \gg \varepsilon_\mathrm{d}$, the rotation axis gradually approaches the magnetic axis ($\chi \to 0$). 
When $|\varepsilon_\mu| \ll \varepsilon_{\rm d}$, the rotation axis of {\it a biaxial} star evolves towards either the deformation axis or the deformation equator, depending on the angle $\alpha$ between the magnetic and symmetry axes. 
Triaxiality in this regime either makes $\pmb\Omega$ aligned with one of the extremal principal axes ($\pmb e_{1,3}$), or keeps it precessing along an attractor orbit. In the former case $\chi$ approaches a final non-zero value, while in the latter case it wobbles on a precessional timescale with a finite amplitude $\Delta\chi$. Finally, in the intermediate regime, $|\varepsilon_\mu|\sim \varepsilon_\mathrm{d}$, we find that the rotation axis either approaches one of the precession poles, which are offset from both the magnetic and the principal axes (see Figure~\ref{fig:lambda_free_precession}), or, as in the weakly magnetized case, enters an attractor state where the long-term evolution approaches a stable limit cycle.  The different evolutionary pathways are summarized in Figure~\ref{fig:evol_cheme}. 

Only in highly magnetized stars ($|\varepsilon_\mu|\gg \varepsilon_{\rm hd}$) $\chi$ evolves according to the standard spin-down model. For the majority of pulsars, we expect the spin and magnetic axes to remain separated essentially at all times.

\subsection{Secular evolution of $\lambda$}
\begin{figure*}
    \centering
    \includegraphics[width=1\textwidth]{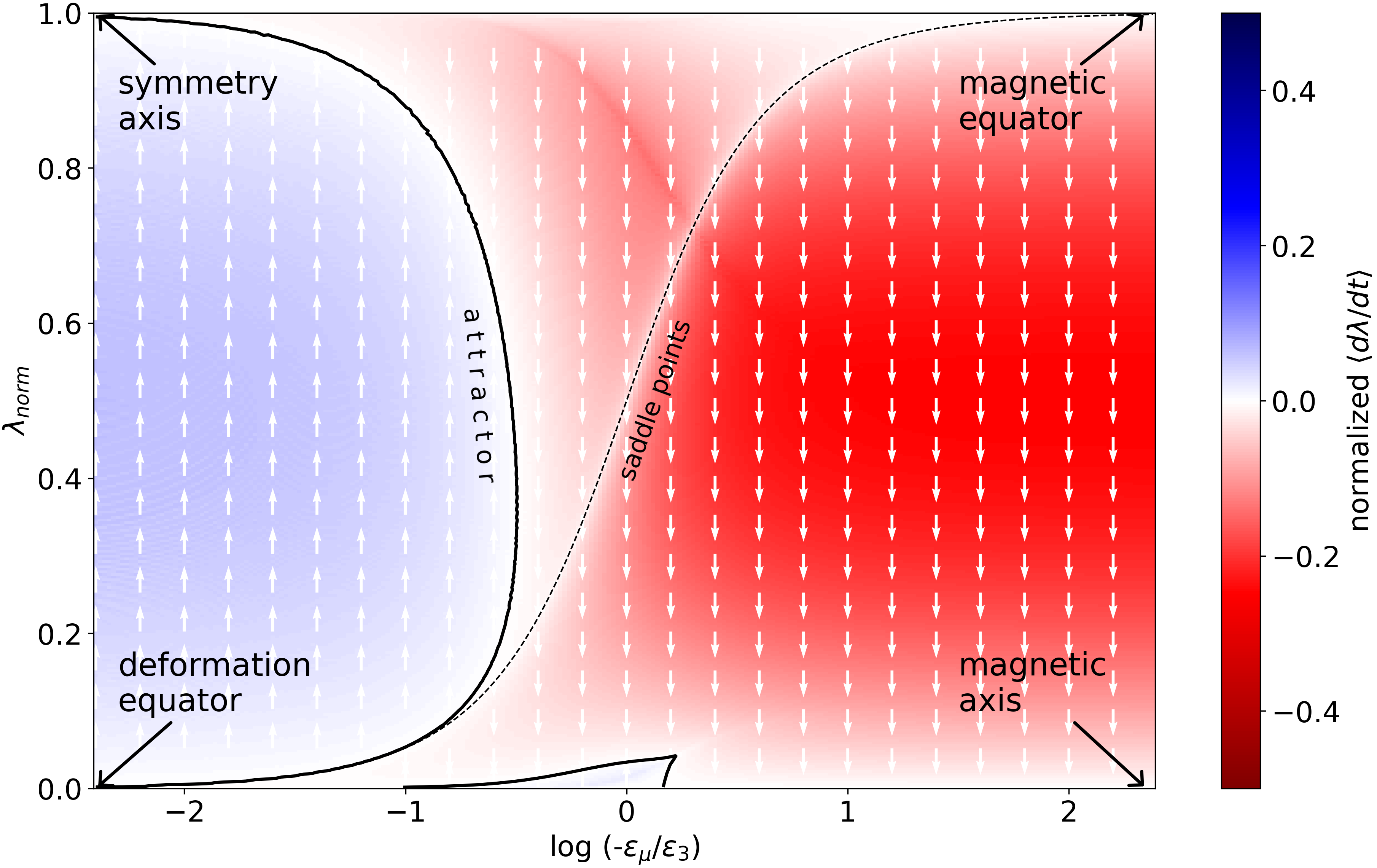}
    \caption{Precession-averaged $\dot{\lambda}$ calculated for an oblate biaxial star with magnetic moment tilted at $45^\circ$ relative to the deformation axis. For $|\varepsilon_\mu| \gg \varepsilon_\mathrm{d}$, $\lambda_\mathrm{norm} = 0$ at the magnetic poles and $1$ at the magnetic equator, while for $|\varepsilon_\mu| \ll \varepsilon_\mathrm{d}$, $\lambda_\mathrm{norm} = 0$ at the deformation equator and $1$ at the deformation (symmetry) axis. White arrows show the direction of evolution of $\lambda$. As seen, in the case of strong radiative precession, the spin axis tends to align with the magnetic axis. On the other hand, when free precession dominates, it aligns with the symmetry axis instead. An attractor exists in a certain range of parameters where $\varepsilon_\mathrm{d} \gtrsim \mathrm{few}\times|\varepsilon_\mu|$, and is marked by the solid black line. The black dashed line marks the locus of saddle points on the map. } 
    \label{fig:oblate45map}
\end{figure*}
%

Upon projecting Euler equations (\ref{eq:Euler_final}) on $\pmb \Omega_\mathrm{p}$ and taking into account the above assumptions ($\varepsilon_\mathrm{d} \ll 1$ and constant, $\varepsilon_\mu = const$, $\pmb m = const$), one obtains an equation for the long-term evolution of $\lambda$ (see Appendix~\ref{app:lambda} for a detailed derivation):
\begin{equation}
     \Omega^2 \frac{\diff\lambda}{\diff t} = \pmb{\Omega}_\mathrm{p} \cdot \pmb n_\chi.
    \label{eq:dot_lambda}
\end{equation}
This equation indicates that $\lambda$ tends to be nearly conserved on the precessional time scales, and its longer-term evolution is governed by the external torque component $n_\chi$, which is responsible for the magnetic angle evolution. 
In a more explicit form,
%
\begin{align}
    \tau_\mathrm{psr}\frac{\diff\lambda}{\diff t} & = \varepsilon_\mu \sin^2\chi\cos^2\chi\, \notag \\
  & \qquad{} + \varepsilon_1 \frac{\Omega_1}{\Omega} \left (m_1\cos\chi - \frac{\Omega_1}{\Omega}\cos^2\chi\right )\, \notag \\
  & \qquad{} + \varepsilon_3 \frac{\Omega_3}{\Omega} \left (m_3\cos\chi - \frac{\Omega_3}{\Omega}\cos^2\chi\right ),
    \label{eq:dot_lambda_components}
    \end{align}
%
where $\tau_\mathrm{psr}$ is the pulsar spin-down time, Eq. (\ref{eq:timescale_pulsar}), and $m_k= \pmb{m}\cdot \pmb{e}_k$ are the projections of the magnetic axis onto the principal axes $\pmb{e}_k$. The three terms on the right-hand side of this equation are related to different precessional motions: the first represents the radiative precession, while the others are related to the contribution of the deformation of the star.  

According to Eq.~(\ref{eq:dot_lambda_components}), $\lambda$ changes on the spin-down timescale $\tau_\mathrm{psr}$, meaning that the spin axis gradually shifts from one precessional orbit to another. 
From Eq.~(\ref{eq:dot_lambda}), one also infers that $\lambda$ always remains constant when the spin and precessional axes become aligned.  Indeed, as $\pmb n_\chi = n_\chi\pmb s_2$ is orthogonal to $\pmb \Omega$ by construction, $\pmb \Omega_\mathrm{p}\cdot \pmb n_\chi = 0$ when $\pmb\Omega_\mathrm{p} \parallel \pmb\Omega$. This condition is fulfilled at the poles of the precession ($\dot\varphi = \dot\theta = 0$), which coincide with points of extrema of $\lambda$. The last statement stems from the relations
\begin{equation}
\frac{\partial\lambda}{\partial\theta} = -\frac{1}{\Omega}\sin\theta\,\dot\varphi, \quad\mbox{and } \quad \frac{\partial\lambda}{\partial\phi} = \frac{1}{\Omega}\sin\theta\,\dot\theta, 
    \label{eq:canonical}
\end{equation}
that hold when $n_\chi =0$.
In other words, Eq.~(\ref{eq:dot_lambda}) (or \ref{eq:dot_lambda_components}) is the generalization of the standard equation $\Omega\dot\chi = n_\chi$, that governs the evolution of $\chi$ for a spherical neutron star. 

The difference in precessional and spin-down time scales allows to reduce the evolution of $\pmb{\Omega}$ to the evolution of $\lambda$. 
The latter is easier to solve with help of averaging within one precessional period (or one precessional orbit):
\begin{equation}
    \langle \dot\lambda \rangle = \frac{1}{T_\mathrm{p}}\oint_{\lambda = const} \dot\lambda \diff t.
\end{equation}
Such averaging can be obtained either analytically or numerically, depending on the system's configuration.

Below, we study in detail the long-term evolution of the magnetic angle of a deformed star by averaging Eq.~(\ref{eq:dot_lambda_components}) and assuming different deformations and magnetic fields.

\subsection{Evolution of a spherical star: the control case}
\label{sec:results:lambda:spherical}

The formalism developed in the previous sections may be used to reproduced the alignment process of a spherical pulsar. 
A spherical star has $\varepsilon_{1} = \varepsilon_{3} = 0$, hence $\lambda = \varepsilon_\mu \cos^2\chi/2$. Eq. (\ref{eq:dot_lambda_components}) then takes the form
\begin{equation}
    \tau_\mathrm{psr}\dot\lambda = \varepsilon_\mu\sin^2\chi\cos^2\chi,
\end{equation}
which, upon using  $\dot\lambda = -\varepsilon_\mu \cos\chi\sin\chi\dot\chi$, reduces to the standard magnetic angle evolution law:
\begin{equation}
    \Omega\dot\chi = n_\mathrm{psr}\sin\chi\cos\chi.
    \label{eq:dotchi_spherical}
\end{equation}
As $\varepsilon_\mu < 0$,  $\lambda$ gradually approaches its minimal value $\lambda_{\min} = \varepsilon_\mu/2$ over the spin-down time. On the star's surface, points of minimal $\lambda$  coincide with the magnetic poles. This process represents standard pulsar magnetic alignment. 

\subsection{Evolution of a biaxial star}
\label{sec:results:lambda:biaxial}
\begin{figure*}
    \centering
    \includegraphics[width=0.48\textwidth]{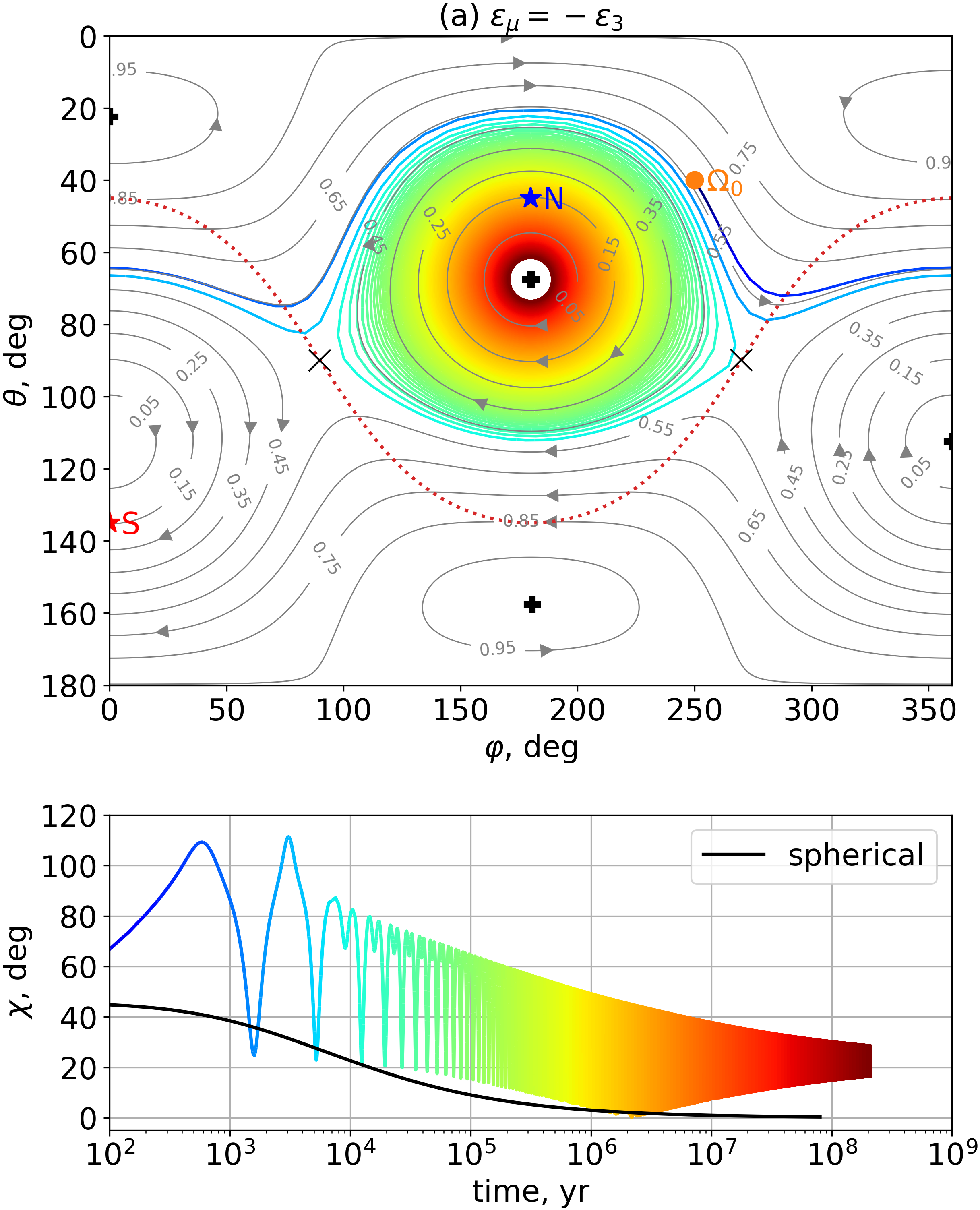}
    \includegraphics[width=0.48\textwidth]{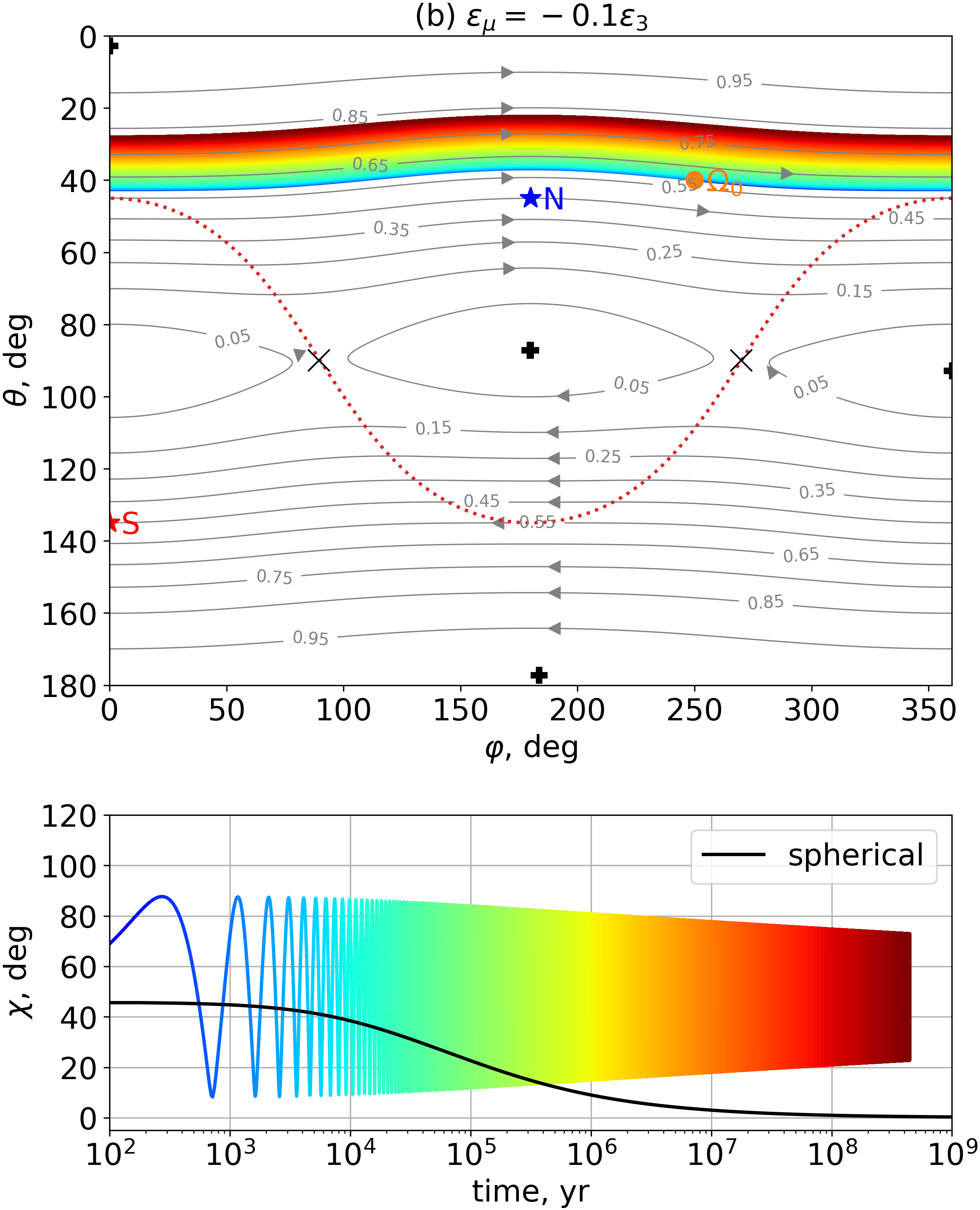}
    \caption{Trajectories of the spin axis for an oblate biaxial star, modeled over several hundred million years, for $\varepsilon_\mu = -\varepsilon_3 = -10^{-12}$ (a), and $\varepsilon_\mu = -0.1 \varepsilon_3 = -10^{-13} $ (b). The gray lines in the upper panels are contours of $\lambda_{\rm norm}$. In both panels, the initial period is $P_0 = 0.02$ s, and the constant magnetic moment is $\mu = 2.25\times 10^{30}$ G cm$^3$. The magnetic axis is tilted relative to the deformation axis $\pmb e_3$ by $45^\circ$. The initial orientation of the spin axis is shown by the orange circle ($\Omega_0$) in the upper panels, while the color of the trajectory represents time. 
    In the upper left panel, the spin axis may be seen approaching one of the precession poles, following a tightly wound spiral trajectory. This precession pole does not coincide with the magnetic pole nor with any of the principal axes. The corresponding evolution of the magnetic angle $\chi$ is exhibited in the bottom left panel; it wobbles significantly and remains nonzero even after millions of years of rotational history. The evolution of $\chi$ for a spherical star with the same parameters is shown for comparison by the solid black line. In panel (b), the spin axis precesses around the symmetry axis and evolves towards the attractor, at $\lambda_\mathrm{\rm norm}^* \approx 0.85$.}
    \label{fig:phase_biaxial_45}
\end{figure*}

Biaxiality is the simplest configuration for which the evolution of the magnetic angle can be affected by free precession. A biaxial star may be oblate or prolate. If Cartesian coordinates corresponding to the axes $\pmb{e}_{1-3}$ are $x$, $y$, and $z$ respectively,
then with our choice of principal axes, the deformation (symmetry) axis of an oblate star ($\Delta I_1=0$) is the $z$-axis and the deformation equator is the circle $x^2+y^2 = a^2$. Here $a$ is the semi-major axis of the spheroid (which is orthogonal to the major principal axis). For a prolate star ($\Delta I_3=0$) the deformation axis is the $x$-axis and the deformation equator is the circle $y^2 +z^2 = b^2$, where $b$ is the semi-minor axis of the spheroid. 

Rotational evolution of a biaxial neutron star has been considered in detail by \citet{goldreich70}, but only for the special case of an almost freely precessing star, where the radiative precessional torque is negligible ($\varepsilon_\mu = 0$). In this case, Eq.(\ref{eq:dot_lambda_components}) can be averaged over the precession period analytically, yielding
\begin{equation}
    \tau_\mathrm{psr} \langle \dot\beta \rangle = \sin\beta\cos\beta \left ( \frac{3}{2}\sin^2\alpha - 1 \right).
\end{equation}
Here, $\beta$ is the angle between $\pmb\Omega$ and the deformation axis, and $\alpha$ is the angle between the magnetic moment and the deformation axis, and angular brackets show averaging over the precessional period. This equation applies both to oblate and prolate stars. It shows that $\pmb\Omega$ tends to align with the symmetry axis when $\sin^2\alpha < 2/3$ ($\alpha \lesssim 55^\circ$ or $\gtrsim 125^\circ$), and to approach the deformation equator otherwise.
\begin{figure*}
    \centering
    \includegraphics[width=\textwidth]{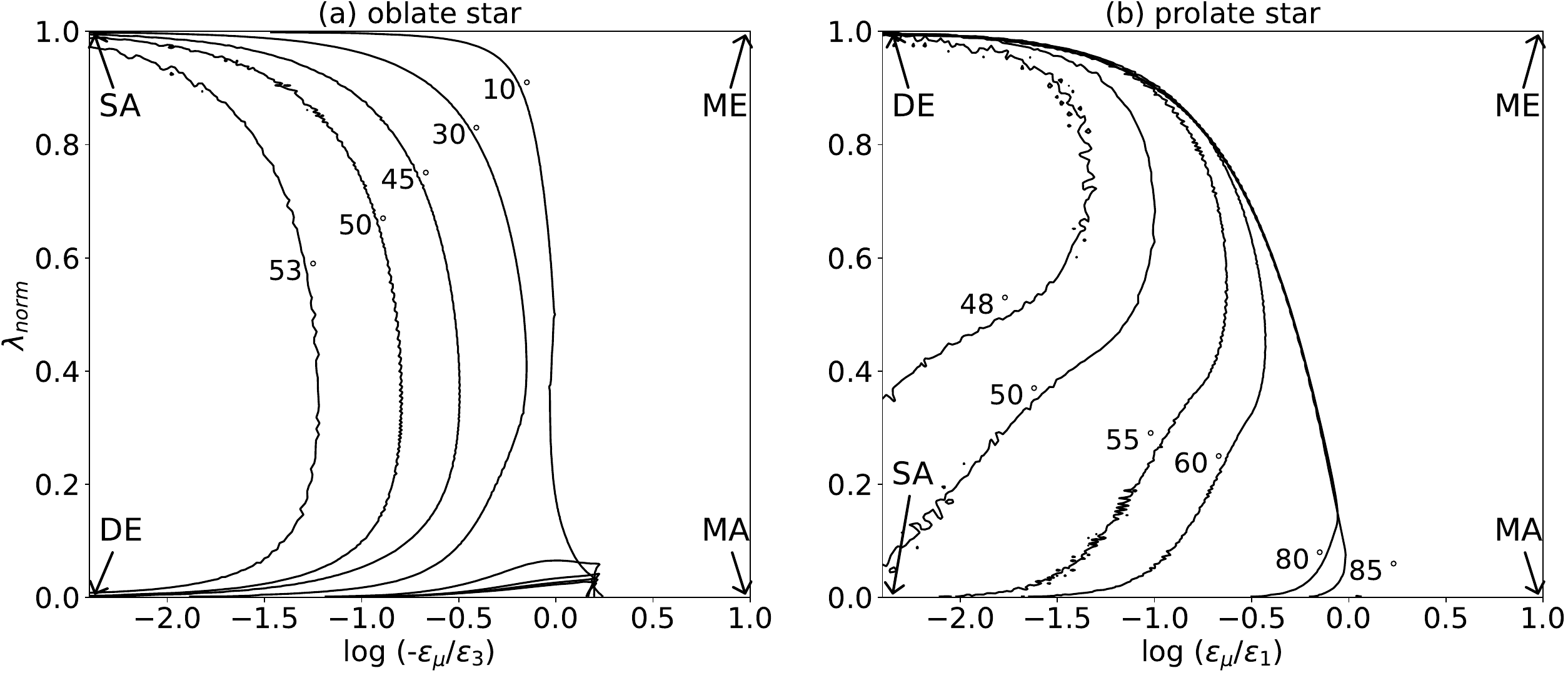}
    \caption{Precessional attractors of a biaxial neutron star, affected by the pulsar torques $n_\chi$ and $n_\mu$. (a) The case of an oblate star ($\varepsilon_3 > 0$, $\varepsilon_1 =0$). The plot is similar to that shown in the Figure~\ref{fig:oblate45map}. The black lines depict attractors in cases when the magnetic axis is tilted at $10^\circ$ to $53^\circ$ relative to the symmetry axis. To the left of these lines, $\lambda$ increases so that the spin axis evolves towards the symmetry axis (SA) when $|\varepsilon_\mu| \ll \varepsilon_\mathrm{d}$. In the case of a strongly magnetized star ($|\varepsilon_\mu| \gg \varepsilon_\mathrm{d}$), $\lambda$ decreases and the spin axis aligns with the magnetic one (MA). ME and DE denote the magnetic and deformation equators, respectively. (b) The case of a prolate star is qualitatively similar to that of an oblate star; however,  the lines marking the attractor have slightly different shapes. The angles in this plot are between the magnetic and deformation axes.} 
    \label{fig:biaxial_complete}
\end{figure*}

In the more general case of $\varepsilon_\mu \neq 0$, the long-term evolution can drive $\lambda$ to one of its extremal points. However, there also exists another evolutionary track for $\lambda$ that approaches a precessional orbit with some specific value $\lambda^*$, $\lambda^* \neq (\lambda_\mathrm{min}, \lambda_\mathrm{max})$, for which
\begin{equation}
    \langle \dot\lambda \rangle = \frac{1}{T_\mathrm{p}}\oint_{\lambda^*} \dot\lambda \diff t = 0.
\end{equation}
This is a state of a precessional attractor (or expeller). We illustrate it in Figure~\ref{fig:oblate45map}, where the map of $\langle \dot\lambda \rangle$ is shown for the case of an oblate biaxial star with magnetic moment inclined at $45^\circ$ to the symmetry axis. The values of $\langle \dot\lambda \rangle$ are in units of $\varepsilon_\mathrm{d}^{-1}\tau_\mathrm{psr}/(\lambda_\mathrm{max}-\lambda_\mathrm{min})$. The vertical axis 
shows $\lambda_{\rm norm}$. The horizontal axis represents the logarithmic ratio between the effective magnetic and real deformations. 
The white arrows on the plot show the direction of the evolution of $\lambda$ at each state. 
The solid black line is the locus of the equilibrium points where $\langle\dot{\lambda}\rangle = 0$. 
This equilibrium may be stable (as the upper part of the large black contour in the figure), which allows one to claim the existence of an attractor orbit on the surface of the NS, or unstable. The dashed line depicts the locus of the saddle points. 
\begin{figure*}
    \centering
    \includegraphics[width=\textwidth]{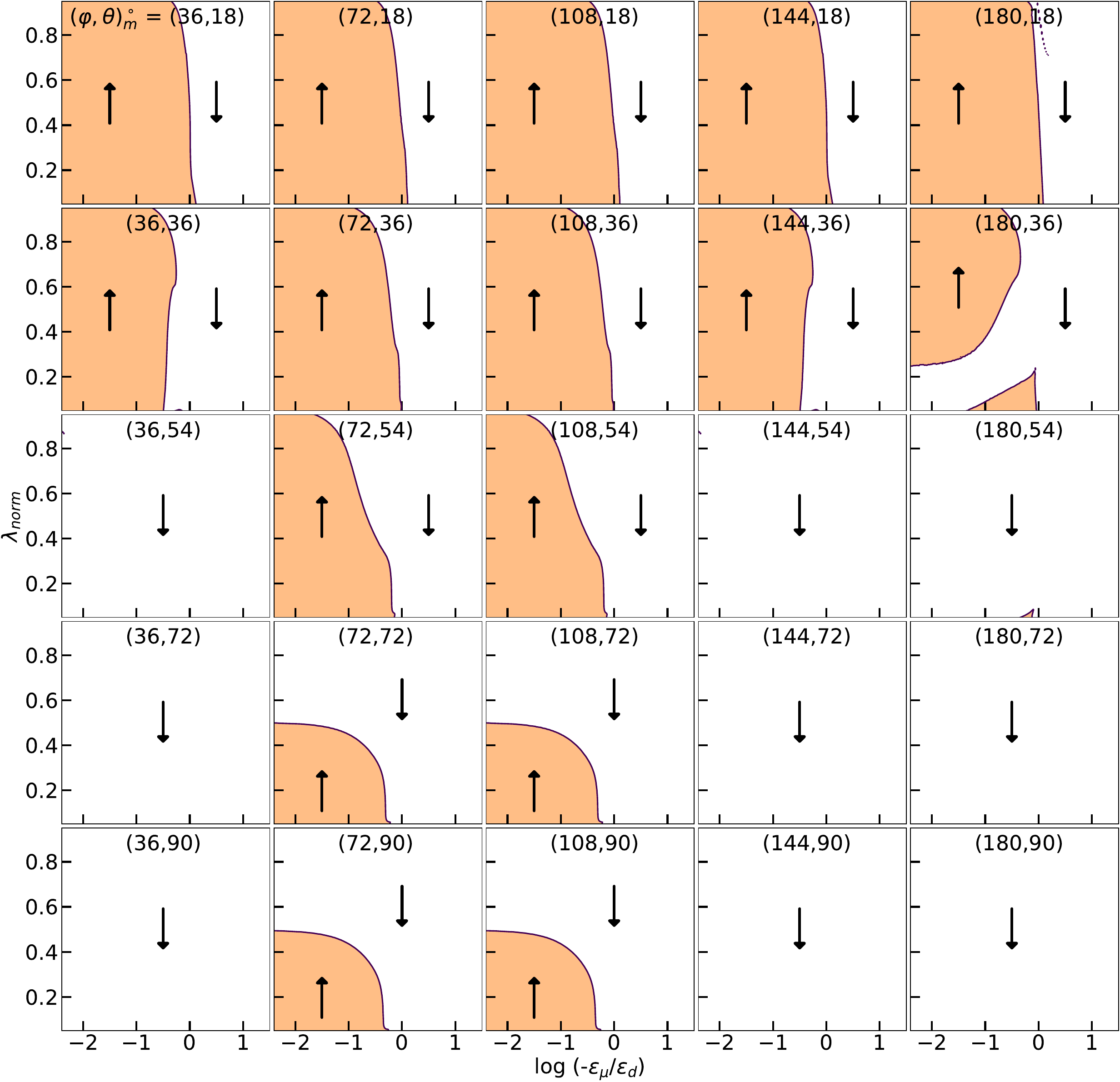}
    \caption{Representation of evolution of a triaxial neutron star with $\varepsilon_1 = -\varepsilon_3$. The plots are similar to those shown in Figs. ~\ref{fig:biaxial_complete}(a,b) and ~\ref{fig:oblate45map}. Shaded regions have $\langle\dot{\lambda}\rangle > 0$. Each plot corresponds to a specific orientation of the magnetic axis, the values of $\varphi_\mathrm{m}$ and $\theta_\mathrm{m}$ are given in degrees in the title of each panel. Due to symmetry with respect the planes $\pmb{e}_1 - \pmb{e}_2$ and $\pmb{e}_1 - \pmb{e}_3$, only the regime $\varphi_\mathrm{m} = 0..180^\circ$ and $\theta_\mathrm{m} = 0..90^\circ$ is shown.} 
    \label{fig:triaxial_evolution}
\end{figure*}

If the star's magnetic field is strong ($|\varepsilon_\mu| \gg \varepsilon_\mathrm{d}$), the precessional poles corresponding to the minimal $\lambda$ value approach the magnetic poles. The mean time derivative $\langle \dot\lambda \rangle$ in this case is always negative, and $\lambda$ gradually evolves towards its minimal value. In other words, the spin axis tends to align with the magnetic axis, as expected from the standard pulsar evolution. 
On the other hand, if $|\varepsilon_\mu| \simeq \varepsilon_\mathrm{d}$, the precession poles no longer coincide with the magnetic poles, and the spin axis tends to align with a new direction, different from both the principal and the magnetic axes. This type of evolution is shown in Figure~\ref{fig:phase_biaxial_45}(a). The upper panels exhibit trajectories of the spin axis in the principal frame. To obtain them, the Euler's equations (\ref{eq:Euler_final}) have been solved numerically, including consistent evolution of the spin period. Color indicates time, in accord with the bottom plot that delineates the course of evolution of the magnetic angle $\chi$. For this particular simulation, we adopted the following parameters: $\varepsilon_3 = 10^{-12}$, initial period $P_0 = 0.02$ s, $\varepsilon_\mu = -\varepsilon_3$, corresponding to $\mu = 2.25\times 10^{30}$\,G\,cm$^{3}$, and a tilt of the magnetic axis relative to $\pmb e_3$ of $\theta_{\rm m} = \alpha = 45^\circ$. The initial orientation of the spin, $\pmb\Omega_0$, is shown by the orange circles in the upper panels of the plot.
Spin axis moves towards one of the precession poles, while the magnetic angle wobbles significantly on the precessional time scale $T_\mathrm{p} \sim 10^3..10^4$ years. 
The black solid lines in the bottom plots show the expected evolution of $\chi$ for a purely spherical neutron star with the same initial parameters. 

In contrast, when free precession dominates, the precession-averaged torque $n_\chi$ drives the spin axis towards the deformation axis instead of the magnetic axis: $\langle \dot\lambda \rangle > 0$ and $\lambda$ evolves to its maximal value. 
Interestingly, an attractor state can be reached for certain values of $\lambda$, when the deformation $\varepsilon_\mathrm{d}$ is only a few times stronger than $|\varepsilon_\mu|$.

The shape of the attractor boundary on this plot depends on the angle $\alpha$ between the magnetic and symmetry axes. In Figure~\ref{fig:biaxial_complete}(a,b), such boundaries are shown for an oblate (a) and prolate (b) star, assuming different values of $\alpha$. These plots are similar to Figure~\ref{fig:oblate45map}: to the left from the attractor boundary, $\lambda$ evolves to its maximal value. The noisy behavior in some of them reflects the precision of numerical calculations undertaken to produce $\dot\lambda$ maps for the considered cases.

We have found that when $\alpha \lesssim 55^\circ$ for an oblate and $\lesssim 35^\circ$ for a prolate star, $\lambda$ continuously decreases in both cases. The more detailed treatment of the NS precessional attractor states is the subject of further research.

\subsection{Evolution of a triaxial star}
\label{sec:results:lambda:triaxial}
\begin{figure*}
    \centering
    \includegraphics[width=0.48\textwidth]{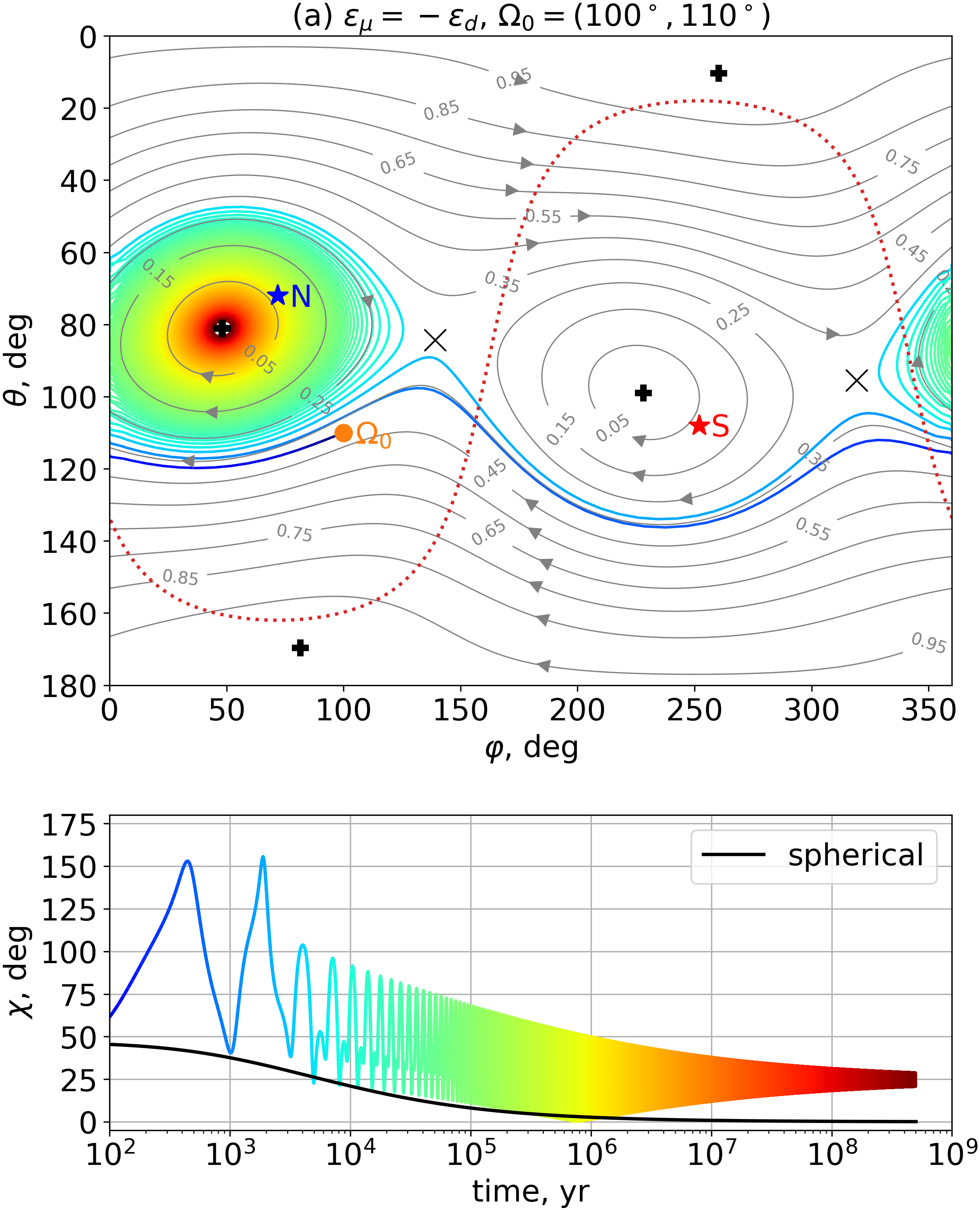}
    \includegraphics[width=0.48\textwidth]{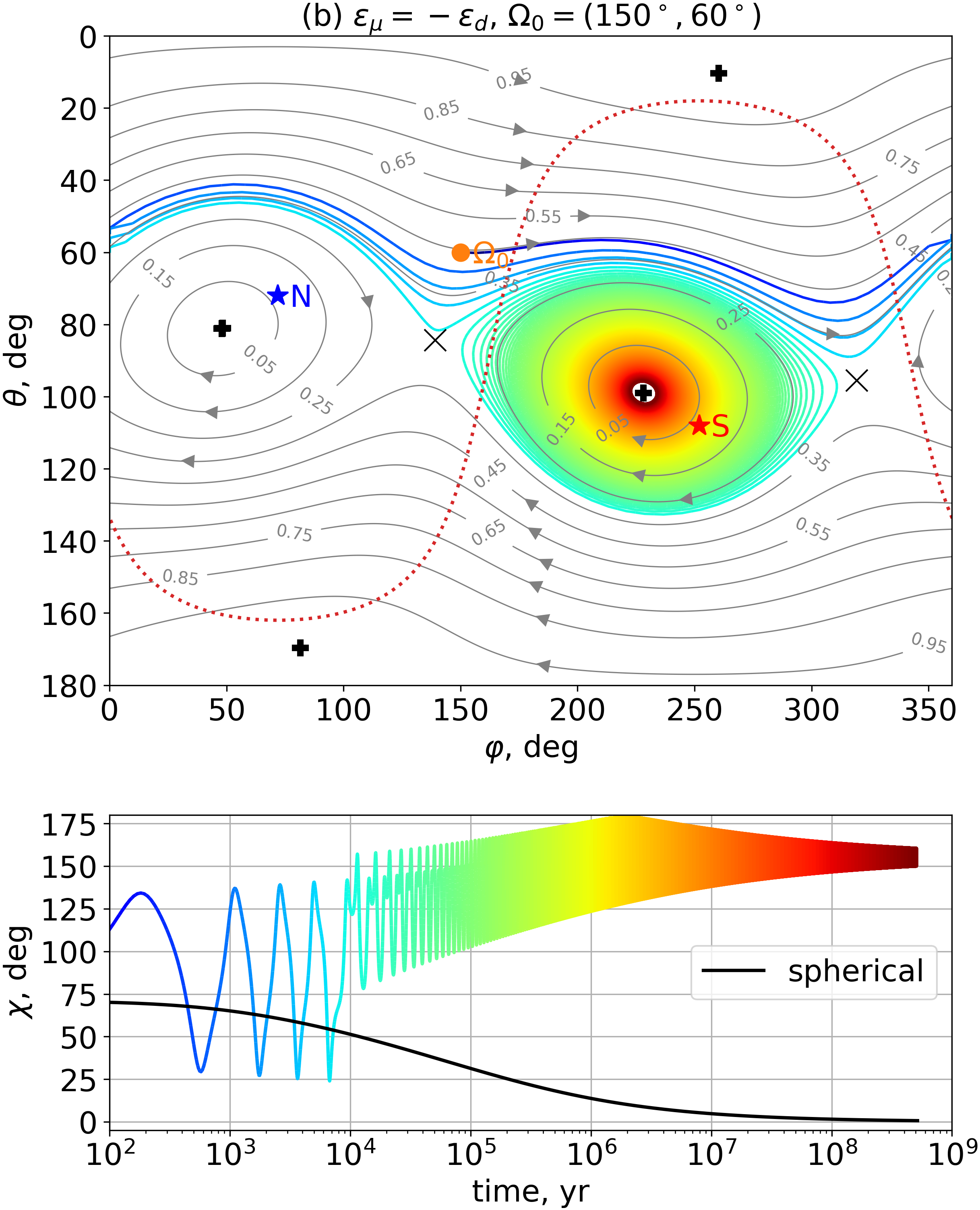}
    \caption{Same as Figure~\ref{fig:phase_biaxial_45}, but for a triaxial star ($\varepsilon_3 = -\varepsilon_1 = 10^{-12}$) with the magnetic axis along the direction $(\varphi_\mathrm{m}, \theta_\mathrm{m}) = (72^\circ,72^\circ)$. The difference between plots (a) and (b) is the initial orientation of the spin axis ($\Omega_0$). In (a), the initial magnetic angle is close to 50$^{\circ}$, while in (b) it is 75$^{\circ}$. As a result of evolution for identical global parameters, the spin axes in the two cases align with {\it different} magnetic poles.}
    \label{fig:phase_triaxial}
\end{figure*}

The evolution of $\lambda$ (and, therefore, the magnetic angle) in the case of a triaxial body is qualitatively similar to that of a magnetized biaxial one.  
It mainly depends on the ratio $|\varepsilon_\mu|/\varepsilon_{\rm d}$ and the initial orientation of the spin axis in the principal frame: for $|\varepsilon_\mu| \gg \varepsilon_{\rm d}$,  $\lambda$ evolves towards the precession poles that nearly coincide with the magnetic poles (Figure \ref{fig:lambda_total_precession}a), meaning that the spin axis gradually approaches the magnetic axis (that is, $\chi\to 0$).  For $|\varepsilon_\mu| \ll \varepsilon_{\rm d}$, the precession and magnetic poles are well separated (Figure \ref{fig:lambda_total_precession}d)
and $\lambda$ evolves towards one of the deformation axes that correspond to extremal moments of inertia (i.e. to the axes $\pmb e_{1,3}$), meaning that the spin and magnetic axes will remain well separated essentially at all times. As a result, the NS evolves toward a non-zero value of magnetic obliquity. For intermediate values of $|\varepsilon_\mu|/\varepsilon_{\rm d}$, $\lambda$ will either evolve toward the precessional axis or reach an attractor state.  As in the previous case, the spin and magnetic axes are expected to remain misaligned. 
In the case of an attractor cycle, the NS at its later stages of evolution also shows oscillations of the magnetic angle $\chi$.

Numerical solutions are shown in Figure~\ref{fig:triaxial_evolution} for a triaxial star with $\varepsilon_3 = -\varepsilon_1 \neq 0$. Each plot represents an evolutionary diagram similar to those shown in Figure~\ref{fig:biaxial_complete} and calculated for several orientations of $\pmb m$: ($\varphi_\mathrm{m}, \theta_\mathrm{m}$) = (36..180$^\circ$, 18..90$^\circ$) at intervals of 36 and 18 degrees, respectively. Due to the symmetry in the values of $\lambda$ within the principal axes, the same behavior is observed in the domain $\varphi_\mathrm{m} > 180^\circ$ and $\theta_\mathrm{m} > 90^\circ$. In particular, the evolutionary plot for $(\varphi_\mathrm{m}, \theta_{m})$ has the same form as for $(\varphi_\mathrm{m} + 180^\circ, 90^\circ-\theta_{\rm m})$).

Some particular evolutionary tracks for $(\theta_\mathrm{m},\varphi_\mathrm{m}) = (72^\circ,72^\circ)$ are also shown in Figure~\ref{fig:phase_triaxial}. In the left and right panels of this figure, the evolution of the same star is presented, with only a difference in the initial orientation of $\pmb{\Omega}$, i.e. different initial magnetic angle. Remarkably, while in both cases $\chi_0$ is less than 90$^\circ$ ($\approx 50$ and $75^\circ$ respectively), the alignment of the spin axis goes towards different magnetic poles, contrary to what is expected for a spherical star.

\subsection{Effects of magnetic field decay}
\label{sec:lambda:decay}
\begin{figure}[ht]
    \centering
    \includegraphics[width=\columnwidth]{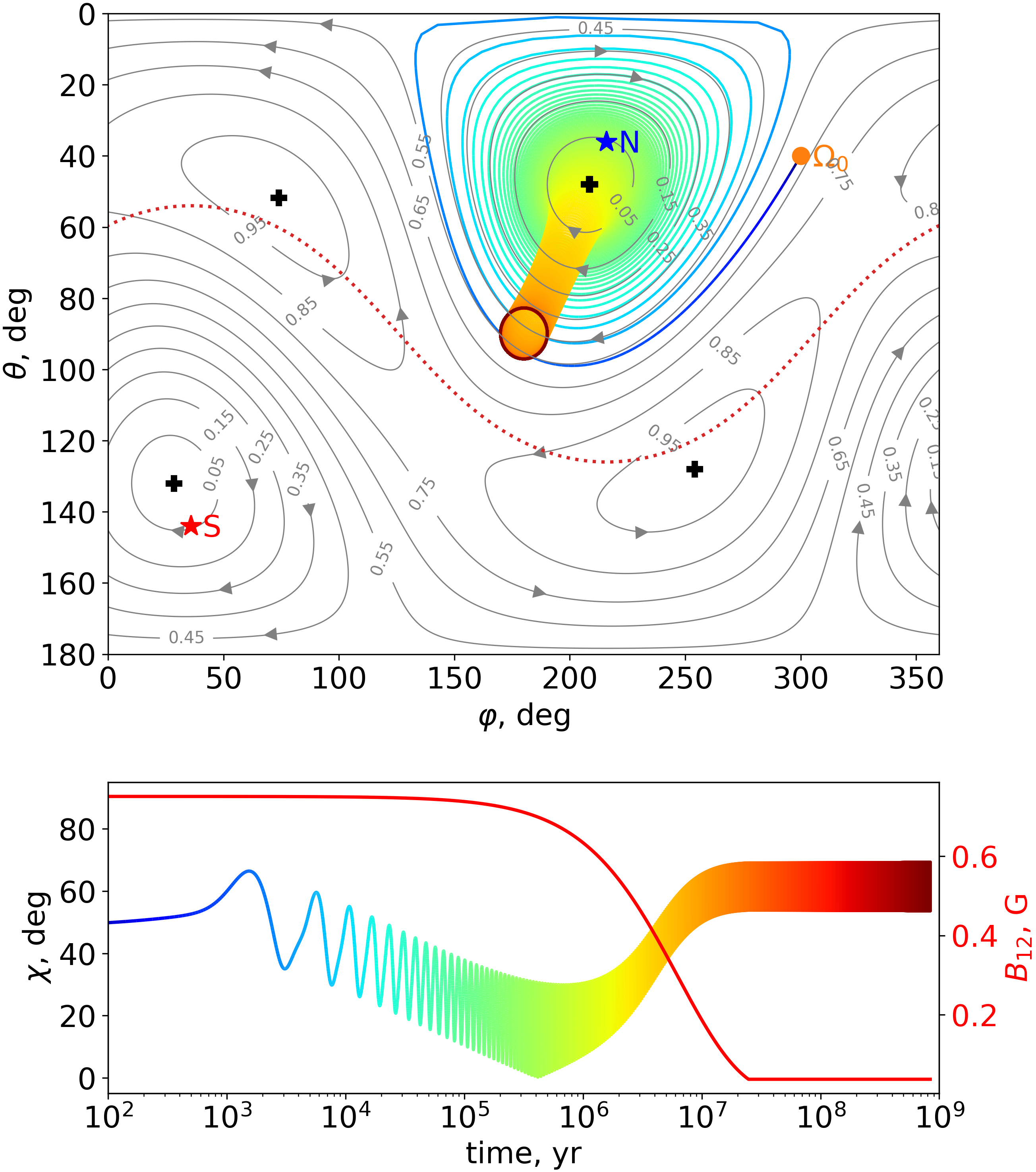}
    \caption{Spin axis trajectory for a triaxial star with magnetic field decay. For this simulation $\varepsilon_{\mu,0} = -3\varepsilon_{d}$ has been adopted. The magnetic field history is shown by the red line in the bottom plot (and the corresponding right vertical axis). During rapid magnetic field decay, the spin axis follows the changing precession axis and ultimately aligns with one of the principal axes. This leads to a gradual increase in the magnetic angle. The radius of the precessional orbit remains practically unaffected during the magnetic field decay, as it occurs much faster than the spin-down time. In the upper panel, the gray contours and the precession poles correspond to the initial, strongly magnetized, configuration.}
    \label{fig:phase_triaxial_decay}
\end{figure}
The rotational evolution of a neutron star, as described above, assumes a vital simplification: the constancy of its magnetic field. However, since it is expected to decay on timescales shorter than spin-down time \citep[e.g.][]{ip21}, the parameter $\varepsilon_\mu \propto \mu^2$ cannot be taken strictly constant. Instead, the precessional orbit of a deformed star, similar to those shown in Figure~\ref{fig:lambda_total_precession}, should evolve with time from larger to smaller values of $|\varepsilon_\mu|$. There are basically three distinct tracks: if the hydrostatic deformation
$\varepsilon_\mathrm{d}$ is always smaller than $|\varepsilon_\mu|$, the evolution will be similar to that of a spherical star. If $|\varepsilon_\mathrm{\mu}|/\varepsilon_\mathrm{d} < 1$, the evolution will be similar to that in the left column of the table in Figure \ref{fig:evol_cheme}. The third possibility is that $|\varepsilon_\mathrm{\mu}|/\varepsilon_\mathrm{d}$  transitions from $> 1$ initially to $< 1$ at a later time. 
We already know the shapes of the precessional orbits for all possible combinations of $\varepsilon_\mu$ and $\varepsilon_{\rm d}$; the question is how $\lambda$ is affected by the evolution of these parameters.

Formally, Eq. (\ref{eq:dot_lambda_components}) must be supplemented in this case with a new term on the right-hand side
\begin{equation}
    \tau_\mathrm{psr}\frac{\diff\lambda}{\diff t} = -\dot\varepsilon_\mu \tau_\mathrm{psr} \cos^2\chi + \tau_\mathrm{psr}\left(\frac{\diff \lambda}{\diff t} \right)_{\varepsilon_\mu = const}.
    \label{eq:dot_lambda_magnetic}
\end{equation}
For a particular set of initial conditions, this evolutionary equation can be integrated numerically if needed. 
However, even without formally solving this equation, the effect of the magnetic field decay can be qualitatively understood. 
Relationships (\ref{eq:canonical}) imply that if $n_\chi = 0$, the scalar $\lambda$ has an analogy to a Hamiltonian of a one-dimensional oscillator described with generalized coordinate $q \equiv \varphi$, momentum $p \equiv \cos\theta$, so that the canonical equations $\partial\lambda/\partial q = - p^\prime$ and $\partial\lambda/\partial p = q^\prime$ are exactly fulfilled. Here, prime denotes the $\partial/(\Omega \partial t)$ derivative. In this analogy, $\lambda$ is the total energy of such an oscillator, while contours plotted in Figures~\ref{fig:lambda_free_precession} and ~\ref{fig:lambda_total_precession} are its phase portraits. The external torque $n_\chi$ can then be imagined as a dissipative term responsible for the energy losses. At the same time, $\varepsilon_\mu$ is one of the parameters of such an oscillator. If $\varepsilon_\mu$ changes on timescales longer than the precession timescale $2\pi/\Omega_\mathrm{p}$, but shorter than the dissipative timescale $\tau_\mathrm{psr}$, then the classical adiabatic invariant,
\begin{equation}
    J = \frac{1}{2\pi} \oint_{\lambda = const} \cos\theta d\varphi,
    \label{eq:decay_invariant}
\end{equation}
should exist. It has a geometric meaning of the surface area enclosed by the $\pmb \Omega$ trajectory in the $\cos\theta - \varphi$ plots. Therefore, during the evolution of the magnetic field, the spin axis will successively follow trajectories that enclose approximately the same area. For a tight and approximately circular orbit, conservation of the enclosed area also implies conservation of its angular radius. 
At later stages, such a trajectory would show a $\chi$ variation amplitude of about the initial diameter of its orbit.

If the magnetic field is strong enough, then the precession poles are close to the magnetic ones. 
Decaying field will make the precession poles move from the magnetic axis towards one of the principal axis of inertia. 
Hence, the average magnetic angle will increase, driving the pulsar towards a nearly orthogonal rotator. We illustrate this effect in Figure~\ref{fig:phase_triaxial_decay}. The contours of constant $\lambda_\mathrm{norm}$ in this plot correspond to the initial state of $\varepsilon_{\mu,0} = -3\varepsilon_\mathrm{d}$ 
(which, taking $\varepsilon_\mathrm{d} = \sqrt{2}\times 10^{-13}$, corresponds to $\mu_0 \approx 1.4\times 10^{30}$\,G\,cm$^3$).

To model the magnetic field evolution, we adopted a phenomenological law based on the numerical results of \citet{agu08}
\begin{equation}
    B = B_0 \frac{\exp(-t/\tau_\mathrm{Ohm})}{1 + (\tau_\mathrm{Ohm}/\tau_\mathrm{Hall}) \left (1 - \exp(-t/\tau_\mathrm{Ohm}) \right)},
    \label{eq:field_decay}
\end{equation}
where $\tau_\mathrm{Ohm} = 10^7$ years is the characteristic time of Ohmic decay and $\tau_\mathrm{Hall} = 10^4\left( 10^{15}{\rm \, G}/B_0\right)$ years is the characteristic time scale of the Hall cascade \citep{Igoshev19, ip21}. We assume that the magnetic field decays in accordance with this relationship until it reaches $0.05B_0$ (after approximately three e-foldings), and then remains constant. This behavior mimics a hypothetical Hall attractor stage \citep{gc14prl, gc14mnras}.

During the period of rapid magnetic field decay (as indicated by the red line in the bottom plot of the figure), the average magnetic angle increases due to the motion of the precession pole. At the same time, the amplitude of the $\chi$ variations, $\Delta\chi \sim $ few degrees, remains nearly constant. This reflects the constancy of the adiabatic invariant (\ref{eq:decay_invariant}).
If the magnetic field evolves faster than the spin-down time scale, it also sets the time for $\chi$ evolution. 
The amplitude of the effect depends on the mutual orientation of the magnetic field and the principal axes. 

We conclude that a decaying magnetic field in a deformed neutron star can give rise to a gradual misalignment of the magnetic and spin axes.  
This is a consequence of the declining radiative precession. Old, weakly magnetized neutron stars tend to precess around one of their principal axes, but the size of the precessional orbit is likely to be defined by the earlier evolution.

\section{Observational implications}
\label{sec:obs}

The precessional time of $\sim (10^{-4}..10^{-2})\tau_\mathrm{psr}$ discussed above is much longer than the typical timespan of neutron star observations ($\lesssim 50$ years). Such a slow precession cannot be observed directly as, e.g., a periodic component in the timing solution. However, 
wobbling of the magnetic angle during the course of evolution can, nonetheless, affect the statistics of pulsar observables. Firstly, it affects the distribution of $\chi$ across the NS population. Secondly, it impacts the observables associated with high-order derivatives of $\Omega$ that are sensitive to $\chi(t)$. In this section, we discuss the basic observational consequences of NS long-period precession, which can explain certain peculiarities in the available data.

\subsection{Statistics of pulsar braking indices}
\label{sec:obs:bi}

Eqs~(\ref{eq:N_Omega_iso}) and (\ref{eq:N_chi_iso}), describing the spin-down of a pulsar, may be used to predict the first and second derivatives of its spin frequency. The derivatives may then be used to calculate the so-called \emph{braking index}\footnote{This dimensionless parameter characterizes the physics of pulsar spin-down. In particular, if one assumes a phenomenological spindown evolution that obeys a simple power-law form, $\dot\Omega = K\Omega^n$, where $K = const$, then $n_\mathrm{obs} = n$.},
\begin{equation}
    n_\mathrm{obs} = \frac{\ddot\Omega \Omega}{\dot\Omega^2}.
\end{equation}
In the most general form (assuming only $I_0 = const$), it is 
\begin{equation}
    n_\mathrm{obs} = 3 - 4\left(\frac{\dot\mu}{\mu} + f(\chi)\dot\chi \right)\tau_\mathrm{ch},
    \label{eq:brakingI_full}
\end{equation}
where $\tau_\mathrm{ch} = -\Omega/2\dot\Omega$, and
\begin{equation}
    f(\chi) = \frac{\sin\chi\cos\chi}{1 + \sin^2\chi}.
    \label{eq:f(chi)}
\end{equation}
For pulsars with constant magnetic field
\begin{equation}
    n_\mathrm{obs} = 3 + 2 \frac{\sin^2\chi\cos^2\chi}{(1+\sin^2\chi)^2} \in 3-3.25,
\end{equation}
which is a clear prediction for the spin-down of a spherical neutron star. Monotonically decaying magnetic field in isolated radiopulsars may lead to larger positive values of $n_\mathrm{obs} \sim $few$\times 10^2$ \citep{ip20}. However, for hundreds of known isolated pulsars, the braking indices reach absolute values of up to $\sim 10^6$ and are negative in approximately half of the cases. This is known as the problem of {\it anomalous braking indices} \citep[e.g.][]{zhang12}.

Unfortunately, estimating $\ddot\Omega$ (and, hence, $n_\mathrm{obs}$) from the observations is difficult. 
While Eqs.~(\ref{eq:N_Omega_iso}) and (\ref{eq:N_chi_iso}) predict a smooth power-law spin-down on the time scale of $\tau_\mathrm{psr}$, the real rotational evolution of the majority of isolated pulsars exhibits irregularities known as ``timing noise'' -- 
quasi-periodic variations on time scales ranging from months to years \citep[e.g.][]{hobbs10}. This phenomenon has been extensively studied over the past several decades, and numerous theories have been proposed to explain it \citep{bransgrove2025, boyn72, ch80, daless95, urama06, hobbs10, zhang12, lyne13, ou16}.  Currently, there is no broad consensus on the specific cause of the timing noise. 

From the perspective of the evolutionary model considered here, it is worth pointing out that the timing noise significantly influences the higher-order derivatives of the rotational frequency, including $\ddot\Omega = \diff^2\Omega/\diff t^2$. Thus, $n_\mathrm{obs}$,
when estimated over a short observational timespan, can lie well outside the predicted range $3 - 3.25$. Although timing noise is thought to be the primary cause of anomalous $n_\mathrm{obs}$, evolutionary effects accumulated over timescales vastly longer than the observational span -- specifically, precession -- can also cause large deviations of $n_\mathrm{obs}$ from the value predicted by the standard model.  It is therefore imperative to disentangle long-term evolutionary effects in observational data from timing noise. In what follows, we offer a method for selecting out pulsars in which $\ddot{\Omega}$ is likely to be 
dominated by long-term evolution. We stress that even in pulsars that exhibit timing noise, long-term evolutionary effects might be important.
\begin{figure}[t]
    \centering
    \includegraphics[width=\columnwidth]{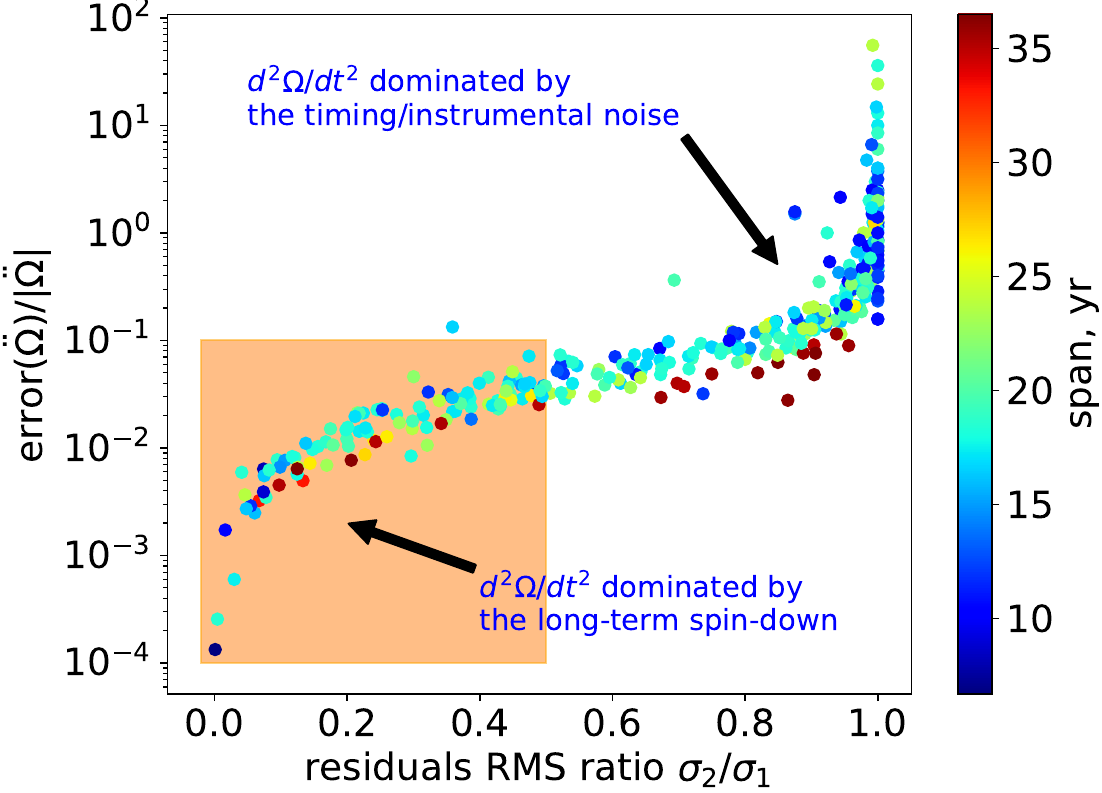}
    \caption{A plot of 366 pulsars whose timing noise properties have been studied in \cite{hobbs10}. The vertical axis shows the relative precision of the measurement of $\ddot{\Omega}$,  
    and the horizontal axis the ratio $\sigma_2/\sigma_1$, where $\sigma_1$ is the timing residuals RMS after subtraction of the quadratic timing model, and $\sigma_2$ is the same after subtraction of the cubic model. 
    The color indicates the time span used for the timing measurement.
    Pulsars within the shaded (orange) area have $|\delta\ddot\Omega/\ddot\Omega| \le 0.1$ and $\sigma_2/\sigma_1 < 0.5$. For them, $\ddot\Omega$ is expected to be dominated by the long-term spin-down evolution rather than timing noise within the measurement duration. The effect
    of long-term evolution on $\ddot{\Omega}$ for the other pulsars in this sample is uncertain.}
    \label{fig:rms}
\end{figure}
\begin{figure*}
    \centering
    \includegraphics[width=1\textwidth]{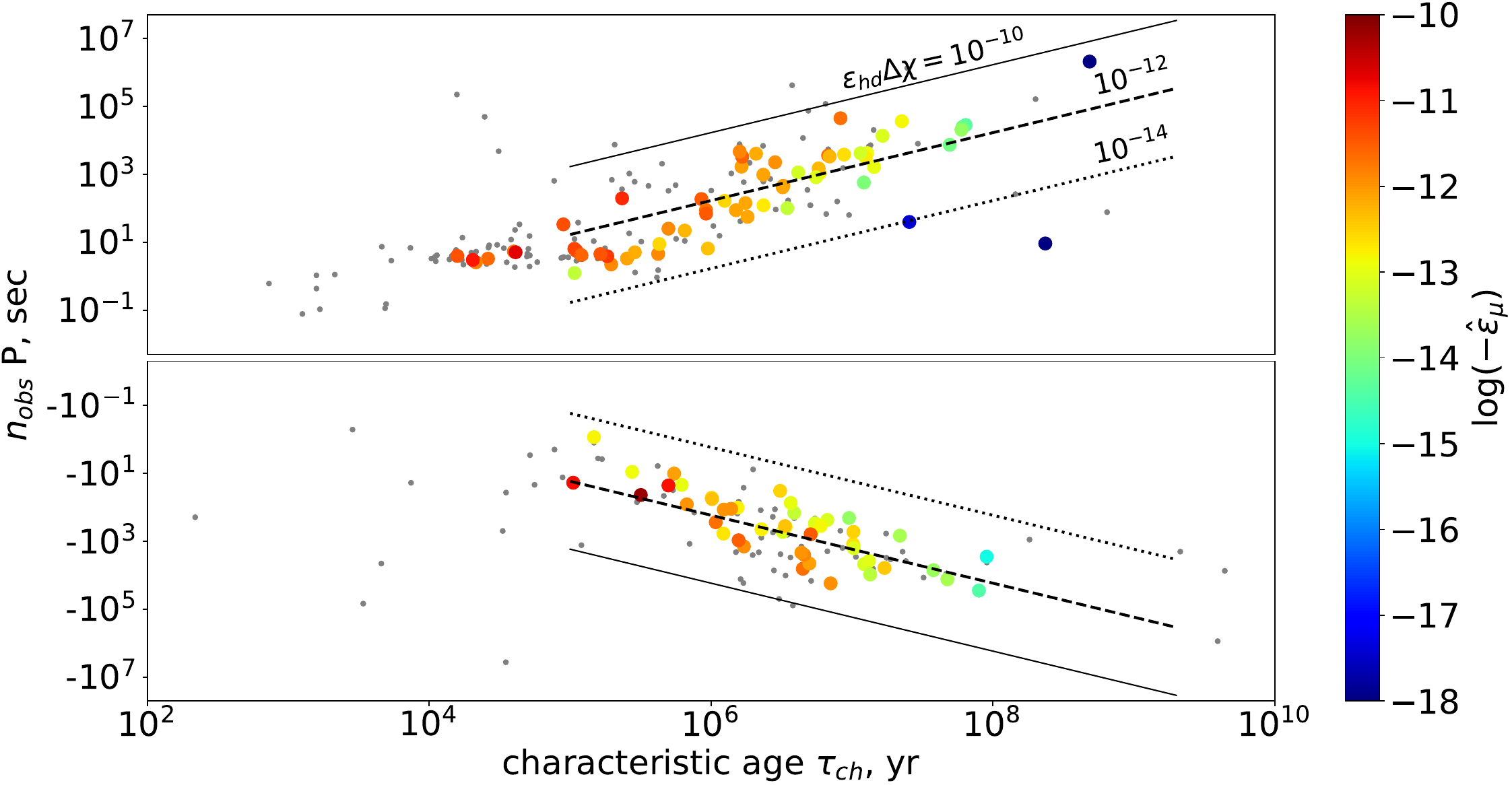}
    \caption{Reduced braking index $n_\mathrm{obs}P$ versus the characteristic age for isolated radiopulsars. The gray dots designate 249 pulsars for which $\ddot\Omega$ is estimated with an accuracy better than 20\% and are stored in the ATNF database. The colored circles depict those 108 pulsars from \citet{hobbs10}, whose $\ddot\Omega$ is likely affected by long-term evolution rather than by timing noise.
    Their color indicates magnetic deformation $\hat{\varepsilon}_\mu$ defined in Eq. (\ref{eq:hat_epsilon_mu}). 
    The lines (solid, dashed, and dotted) show the expected behavior of the braking index for hydrostatically deformed neutron stars (Eq. \ref{eq:averaged_bi}) 
    with deformation parameters $\varepsilon_\mathrm{d}\Delta \chi = 10^{-10}$, $10^{-12}$, and $10^{-14}$.} 
    \label{fig:ntau}
\end{figure*}
Computationally, $\ddot\Omega$ represents the cubic term in the pulsar rotational phase expansion,
\begin{equation}
    \psi(t_0+\Delta t)=\psi_0+\Omega \Delta t+\frac{1}{2}\dot\Omega \Delta t^2 + \frac{1}{6}\ddot\Omega \Delta t^3+...\, .
    \label{eq:phase_poly}
\end{equation}
The RMS of the phase residuals obtained after subtraction of a polynomial model (\ref{eq:phase_poly}) from the observation is a convenient way to estimate the contribution of timing noise to pulsar rotation. 
Following \citet{hobbs10}, let $\sigma_1$ denote the RMS residuals after subtracting the quadratic polynomial (i.e., estimating $\Omega$ and $\dot\Omega$ only), and $\sigma_2$ is the same for the cubic model. If subtraction of the cubic model significantly reduces the RMS of the residuals ($\sigma_2 < \sigma_1$), then the resultant $\ddot\Omega$ is likely dominated by the long-term component of its spindown and does not reflect the timing noise within the observational span. 
In this case, one might also expect that $\ddot\Omega$ is estimated with high statistical precision, so the relative error $|\delta\ddot\Omega/\ddot\Omega| \ll 1$. 

We have analyzed 366 isolated pulsars from the sample of \citet{hobbs10}. 
From them, we have selected 108 objects for which $|\delta\ddot\Omega/\ddot\Omega| \le 0.1$ and $\sigma_2/\sigma_1 < 0.5$.
The selection condition is shown in Figure~\ref{fig:rms}  as a shaded area. 
Also, all the pulsars of the sample are shown in the figure as dots with the color encoding the length of the available observational timespan.

We have calculated the braking indices and characteristic ages for the 108 selected pulsars (see Figure~\ref{fig:ntau}). The braking indices of all of them deviate substantially from the standard prediction, and are even negative in approximately half of the cases. This is seen in Figure~\ref{fig:ntau}, where the reduced braking index,
\begin{equation}
    n_\mathrm{obs} P = \frac{2\pi \ddot\Omega}{\dot\Omega^2},
    \label{eq:reduced_bi}
\end{equation}
is plotted against the characteristic age, $\tau_\mathrm{ch}$, for isolated radio pulsars. 
The gray dots there represent the 249 pulsars for which $\ddot\Omega$ was estimated with a formal accuracy of at least 20\%, and stored in the ATNF compilation database.   
The colored circles show the 108 pulsars from \citet{hobbs10} for which, according to our analysis, the braking index is most likely dominated by long-term spin-down evolution. The color indicates the values of $\hat{\varepsilon}_\mu$ estimated from Eq. (\ref{eq:hat_epsilon_mu}). The cause of the anomalous braking index in the remaining (uncolored) pulsars in the ATNF database is unclear at present, and further analysis is needed to assess it.
A strong positive correlation between the characteristic age and reduced braking index is seen for the selected sample of 108 pulsars. 
It has previously been proposed that this kind of anomalous behavior (negative and positive braking indices correlated with the estimated age) is consistent with a quasi-cyclic component in pulsar spin-down acting on time scales of hundreds to thousands of years. 
It has been modeled by adding to the spin-down law an ad hoc oscillatory term, whose underlying physics was either unspecified \citep[e.g][]{urama06, bbk07, bbk12, zhang12} or inconsistent with the actually observed values \citep[e.g][]{pons12, ou16}. 
Moreover, in those previous works, pulsars were not sorted out by their timing behavior, as in our analysis, and it has not been checked whether a long-term component of their spin-down dominates their timing residuals. In what follows, we demonstrate that the observed $n_\mathrm{obs}P-\tau_\mathrm{ch}$ correlation can be {\it quantitatively} reproduced by precession of a population of sufficiently deformed ($\varepsilon_{\rm d} \gtrsim |\varepsilon_\mu|$) NSs.

The wobbling of the spin axis relative to the magnetic axis can significantly affect the observed timing parameters of radio pulsars, since the pulsar spin-down rate,
\begin{equation}
    \dot\Omega \approx n_\Omega = -\frac{\mu^2\Omega^3}{I_0 c^3} \left(1+ \sin^2\chi \right),
\end{equation}
depends on the instantaneous magnetic angle $\chi$. 
This formula for $\dot\Omega(\chi)$ has been tested against observational data of pulsar spin down for nearly orthogonal and nearly aligned rotators \citep{bb23}. Although large variations in $\Omega$ itself and $\dot\Omega$ are unlikely to be produced by such a mechanism,
the second derivative of $\Omega$ could be significantly altered, thereby dramatically changing the observed values of the braking index.
%
\begin{figure*}
    \centering
    \includegraphics[width=\textwidth]{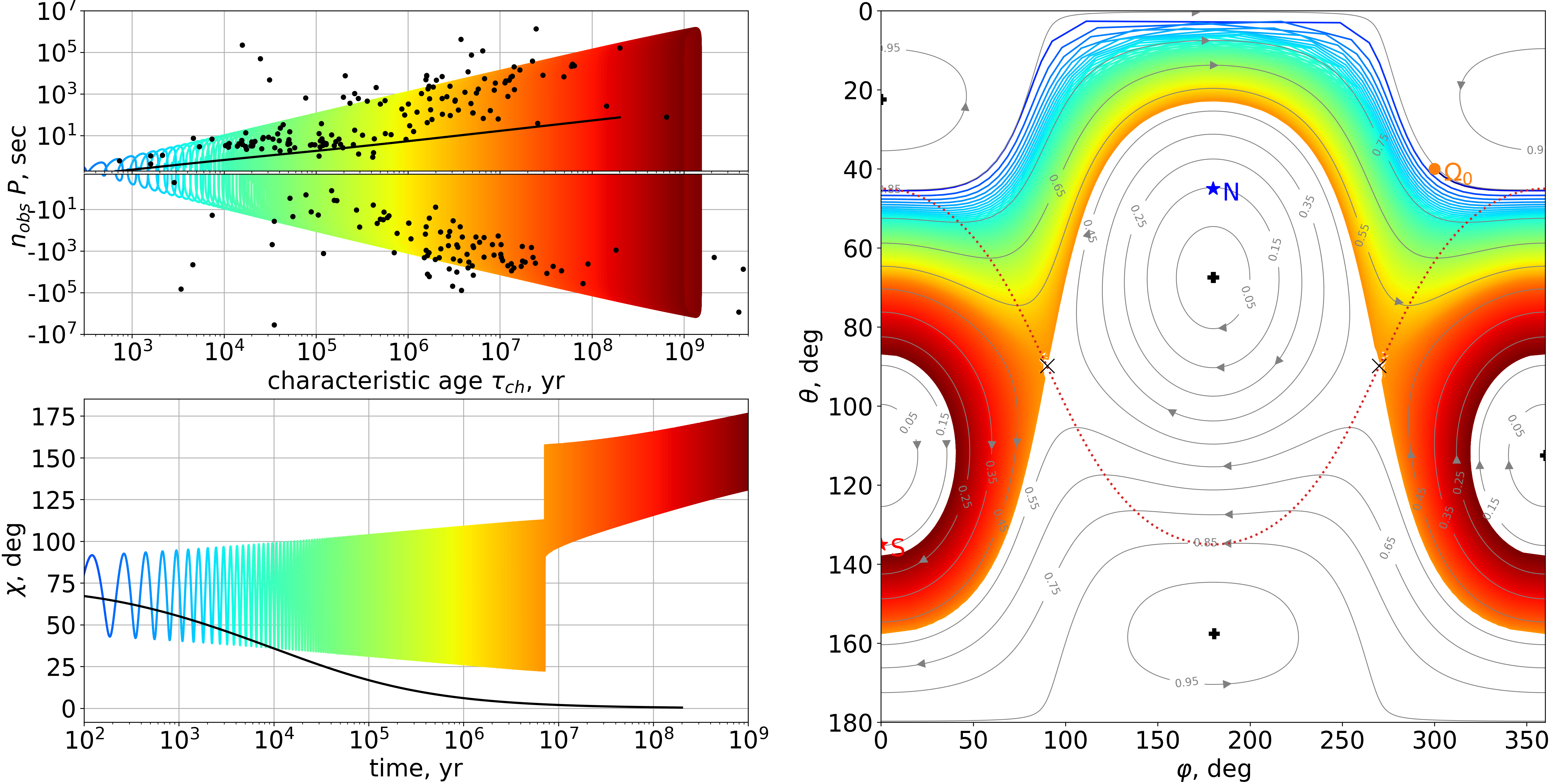}
    \caption{Example of the evolution of $n_{\rm obs}\cdot P$ (upper left panel) of a biaxial oblate star with $\theta_\mathrm{m} = 45^\circ$,
 $\varepsilon_\mu = -\varepsilon_3 = -10^{-11}$, and fixed magnetic dipole. 
 The color indicates the actual age of the pulsar, given in the $\chi(t)$ plot (bottom left panel). 
 The black dots on the $n_\mathrm{obs}P-\tau_\mathrm{ch}$ plot (upper left panel) depict the 249 pulsars from the ATNF sample with known braking indices shown in Figure~\ref{fig:ntau}.}
    \label{fig:bi_modeling}
\end{figure*}

To illustrate the effect of precessional motion on $n_{\rm obs}$, let us assume that the magnetic angle is subject to pure harmonic oscillations with average $\chi_0$, amplitude $\Delta\chi$ and constant precessional frequency $\Omega_\mathrm{p}$. These assumptions are reasonable when considering the evolution of a pulsar over a few precessional cycles. Let us assume that the precessional oscillations of $\chi$ take place on a time scale shorter than both the secular change of its mean value and the magnetic field decay time, $\mu/\dot\mu$. Therefore, they will likely dominate the value of $n_\mathrm{obs}$, making it vastly different from $\sim 3$ and even negative if $\Delta\chi$ is high enough. 
From Eq.~(\ref{eq:brakingI_full}), one gets that the braking index will oscillate over a period $T_\mathrm{p} = 2\pi/\Omega_\mathrm{p}$, with an amplitude of
\begin{equation}
    \Delta n_\mathrm{obs} \approx 4|f(\chi_0)|\,\Omega_\mathrm{p}\,\Delta\chi\,\tau_\mathrm{ch}.
\end{equation}
Since oscillations of $\chi$ are significant only when deformation cannot be neglected in the precessional motion, we can safely adopt $\Omega_\mathrm{p} \approx \varepsilon_\mathrm{d}\Omega$. If $\chi_0$ is distributed isotropically among pulsars of a given characteristic age,  $\langle 4|f(\chi_0)|\rangle = 4-\pi$. Therefore, one finds:
\begin{equation}
    \langle \Delta n_\mathrm{obs} \cdot P \rangle \approx 1.7\times 10^{3}\, \varepsilon_{\mathrm{d},-12}\, \Delta \chi \,\tau_{\mathrm{ch},7}\quad \text{sec},
    \label{eq:averaged_bi}
\end{equation}
where the spin period $P$ is measured in seconds, $\tau_{\rm ch,7} = \tau_{\rm ch}/(10^7\, \text{yr})$, and the deformation parameter  $\varepsilon_\mathrm{d}$ is normalized to $10^{-12}$. 

The lines of $\langle \Delta n_\mathrm{obs} P\rangle$ are overplotted in Figure~\ref{fig:ntau} assuming $\Delta\chi = 1$ and $\varepsilon_\mathrm{d} = 10^{-10}$, $10^{-12}$ and $10^{-14}$. 
These lines are the envelopes of the oscillation process: for each characteristic age value, the model predicts a broad distribution in reduced braking index limited by the values given by Eq.~(\ref{eq:averaged_bi}). 
The correlation in the data (and the braking indices themselves) is in good agreement with that predicted by Eq.~(\ref{eq:averaged_bi}) for reasonable values of $\varepsilon_\mathrm{d}$.
Note that Eq.~(\ref{eq:averaged_bi}) merely sets a lower limit on $\varepsilon_\mathrm{d}$ for individual pulsars because it lacks information on their precessional phases during the observations.  
The requirement that
$|\varepsilon_\mu|/\varepsilon_\mathrm{d} \lesssim 1$, inherent in our model assumptions, is satisfied for this sample by the
two lines (one for positive values and one for negative values of $n_\mathrm{obs} P$) that correspond to $\varepsilon_\mathrm{d}\Delta\chi \sim 10^{-10}$.

In Figure~\ref{fig:bi_modeling}, the results shown in Figure~\ref{fig:ntau} are compared to a simulation
of a biaxial oblate star with $\varepsilon_\mu = -\varepsilon_3 = -10^{-11}$ ($\mu \approx 7\times 10^{30}$ G cm$^3$), and $P_0 = 0.02$ s.
The reduced braking index of this synthetic pulsar oscillates on the precessional time scale with a linearly increasing amplitude (upper left panel). The gray dots depict the same 249 pulsars shown in Figure \ref{fig:ntau}. 
The trend seen in Figure~\ref{fig:ntau}  can be reproduced by a large enough population of such pulsars observed at different times and phases of their evolution. Adding more complexity (e.g., triaxiality,  magnetic field decay, considering a distribution in initial parameters) will merely increase the dispersion of breaking indices. 

The salient conclusion is that the distribution of the reduced braking indices of isolated pulsars over the interval $\sim -10^{6}..10^{6}$ and the approximate linear scaling with the pulsar age may be interpreted as the result of long-term periodic variations in $\chi$ due to free precession caused by NS deformation of magnitude $\varepsilon_\mathrm{d} \sim 10^{-12}..10^{-10}$. 


\subsection{Apparent distribution of magnetic angles}
\begin{figure*}
    \centering
    \includegraphics[width=\textwidth]{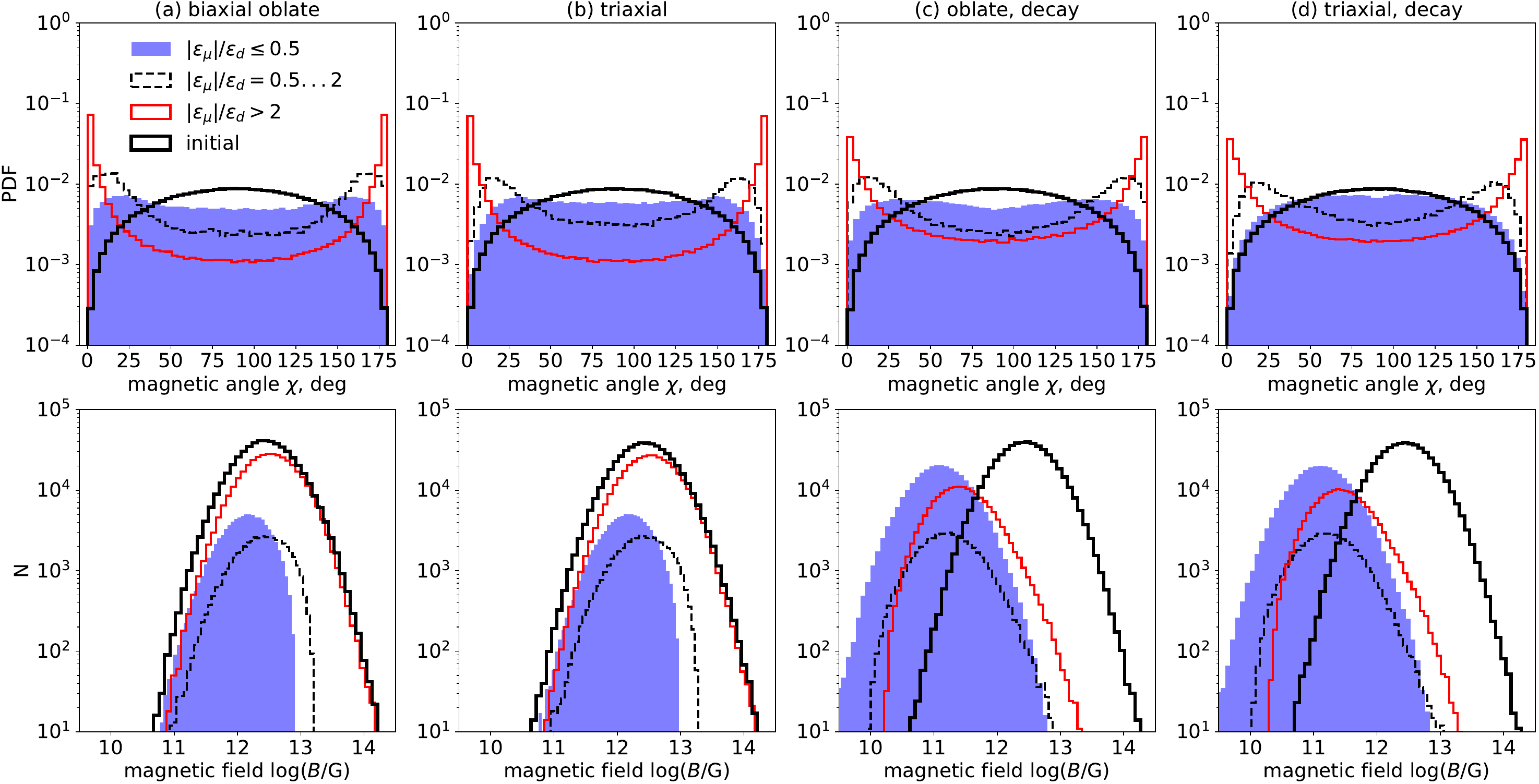}
    \caption{{\it Top row}: Synthetic distributions of magnetic angles of pulsars with ages uniformly distributed up to $10^8$ yrs. All distributions in this row are normalized to unity. 
    Initial spin periods and magnetic field were taken from \citep{Igoshev22}. The distribution of the initial spin and magnetic axes were taken to be isotropic in the principal frame. The values of $\log \varepsilon_3$ were drawn from a uniform distribution in the interval -16..-10. In the simulations of triaxial stars (cases b and d) $\varepsilon_1 = -\varepsilon_3$ was adopted. The plot shows that the magnetic angle distribution of highly magnetized stars (whose precession is weakly affected by deformation) is highly non-uniform with an excess of aligned rotators. Weakly magnetized stars show a uniform or isotropic distribution in $\chi$. {\it Bottom row}: Histograms of the magnetic fields of the same pulsars. They are not normalized, hence the areas under the histograms reflect fractions of pulsars within a particular range of $|\varepsilon_\mu|/\varepsilon_\mathrm{d}$. If the field decays with time, a larger fraction of pulsars have a low $|\varepsilon_\mu|/\varepsilon_\mathrm{d}$ ratio -- see plots (c) and (d). Such pulsars then will dominate the distribution of $\chi$, which becomes more isotropic.}
    \label{fig:chi_distribs}
\end{figure*}

As stated above, the magnetic angle is one of the key observables of radio pulsars, used to study the structure of NS magnetospheres and their evolution. Although $\chi$ cannot be measured directly, several methods have been developed for its estimation, including the analysis of pulse width and shape, as well as polarization data \citep[e.g.][]{lyne88, rankin90, nikitina11, johnson14, FAST_polarization23}. It has been found that the magnetic angles of NSs span a wide range, from nearly aligned to nearly orthogonal. This is true for both classical isolated pulsars \citep[e.g.][]{tm98}, and millisecond pulsars evolving in binary systems \citep{chen98, johnson14}.

In the absence of free precession, Eq. (\ref{eq:N_chi_iso}) implies asymptotic alignment of the magnetic and spin axes, whereby $\chi \to 0$ over a time of $\tau_\mathrm{psr} \sim 10^{6}..10^{7}$ years for a typical NS (see also Eq.~\ref{eq:timescale_pulsar}). 
This timescale is comparable to the maximal age of a classical pulsar near its death line (see Figure~\ref{fig:ppdot}).
Ignoring the effects of free precession, one might naively expect a non-uniform distribution in magnetic angles with an excess of nearly aligned rotators and a lack of orthogonal ones. Such an asymmetry may be even more prominent, since the pulsar life time depends on the magnetic angle: nearly orthogonal rotators switch off more rapidly \citep{bes93}.

Among more than 4000 known pulsars, $\sim$50 orthogonal and  $\sim 25$ aligned rotators have been detected \citep{bb23}. 
\citet{Novoselov2020} argued that the fraction of observed orthogonal rotators with interpulses cannot be reproduced by the standard model (Eqs~\ref{eq:N_Omega_iso} and \ref{eq:N_chi_iso}), thus providing evidence for increasing magnetic obliquity during the course of spin evolution (see also \citealt{Kniazev25}). 
However, if a neutron star is even slightly deformed, the magnetic angle evolves in a complex way that can significantly impact its apparent distribution across the pulsar population. 

To illustrate this effect, we generated synthetic distributions of magnetic angles by solving the rotational evolution of deformed stars for $\sim 5\times 10^5$ sets of initial parameters. 
In order to obtain an evolutionary track, we randomly draw the initial period $P_0$ and magnetic field $B_0$ from  log-normal distributions with averages $\langle \log (P_{0}/\mathrm{s}) \rangle = -1.04$ and $\langle \log (B_0/\mathrm{G})\rangle = 12.44$, and standard deviations $\sigma[\log(P_0/\mathrm{s})] = 0.53$ and $\sigma[\log (B_0/\mathrm{G})] = 0.44$, respectively \citep{Igoshev22}. 
We assign an age for each synthetic pulsar in our sample by randomly drawing a number from a uniform distribution in the range $0 - 10^8$ yrs. Each pulsar is then evolved up to its assigned age. The initial orientations of the spin and magnetic axes of the pulsars, as measured in the principal frame, were taken to be isotropic and independent. 
The value of $\log\varepsilon_3$ was randomly drawn from a uniform distribution in the interval $[-16,-10]$.

The entire sample is divided into four subclasses: (a) biaxial oblate ($\epsilon_1=0$) stars with a fixed magnetic moment; (b) triaxial stars with $\varepsilon_1 = -\varepsilon_3$ and fixed magnetic moment; (c) biaxial oblate stars with a decaying magnetic field; (d)  triaxial stars as in (b), but with a decaying magnetic field. For the magnetic field evolution in the last two cases we adopted Eq. (\ref{eq:field_decay}).

The resultant distributions are shown in Figure~\ref{fig:chi_distribs}. 
Each panel shows the distributions of three different subsets of the same subclass, selected by cuts of $\varepsilon_\mu/\varepsilon_{\rm d}$ as indicated, along with the initial 
distribution (shown by the black solid lines). 
The red solid line represents highly magnetized objects with $|\varepsilon_\mu|/\varepsilon_\mathrm{d} > 2$, for which free precession does not affect the evolution of the magnetic angle significantly, and a gradual alignment of the spin and magnetic axes is expected. The distribution exhibits clear peaks at $\chi = 0$ and $180$ degrees, in agreement with expectations.  
The result is insensitive to the choice of cutoff, $|\varepsilon_\mu|/\varepsilon_\mathrm{d}$, as long as it is larger than $2$.
In contrast,  weakly magnetized stars with $|\varepsilon_\mu|/\varepsilon_\mathrm{d} \le 0.5$  (depicted by the blue area) exhibit a nearly uniform $\chi$ distribution for every choice of initial parameters, except for case (d), where it is closer to isotropic (that is, uniform
in $\cos\chi$). The intermediate case ($|\varepsilon_\mu|/\varepsilon_\mathrm{d} = 0.5..2$) is shown by black dashed lines on these plots.

Ultimately, the resultant $\chi$ distribution of real pulsars would depend on the distribution of the parameter $|\varepsilon_\mu|/\varepsilon_\mathrm{d}$ among the entire population. If $|\varepsilon_\mu|/\varepsilon_\mathrm{d} < 0.5$
preferentially, the distribution should be close to uniform in $\chi$ or isotropic. But even if initially most pulsars 
have $|\varepsilon_\mu|/\varepsilon_\mathrm{d} > 2$, magnetic field decay can drive this ratio to low enough values over times
shorter than the spin down time, as illustrated in Fig. \ref{fig:chi_distribs}(c,d), resulting in an isotropic $\chi$ distribution.

Therefore, we conclude that free precession can account for the observed excess (with respect to the distribution expected for spherical stars) of nearly orthogonal rotators and the lack of nearly aligned ones, provided $|\varepsilon_\mu| < \varepsilon_\mathrm{d} $
for a large enough fraction of evolved pulsars.

\section{Discussion}
\subsection{Braking indices: precession or timing noise?}
\label{sec:discuss:noise}

Although our approach is successful in reproducing the statistics of pulsar braking indices, it does not provide a conclusive proof for the existence of mechanical precession. 
As mentioned above, the observed pulsar spin-down is contaminated by the timing noise. In particular, the residuals obtained by subtracting a basic polynomial model (Eq.~\ref{eq:phase_poly}) from the observational data represent a noise-like process. It can be treated as a quasi-random variation in either rotational phase, spin period, torque, etc. The power spectral density (PSD)
of the residuals $\Delta t = t_\mathrm{obs} - t_\mathrm{model}$ in the pulse times of arrivals (TOAs) shows its consistency with the power law $PSD\propto f^{-p}$, where $f$ is frequency and the index $p$ spans the range $\sim 0$ to $\sim 7$ over the pulsar population \citep[e.g][]{1997JApA...18....5D, 1999MNRAS.306..207B, 2024Univ...10..105Y}. In principle, a strong 
long-term (thousands of years and longer) component of the timing noise could exist and, therefore, affect the estimated $\ddot\Omega$.
\begin{figure}[t]
    \centering
    \includegraphics[width=\columnwidth]{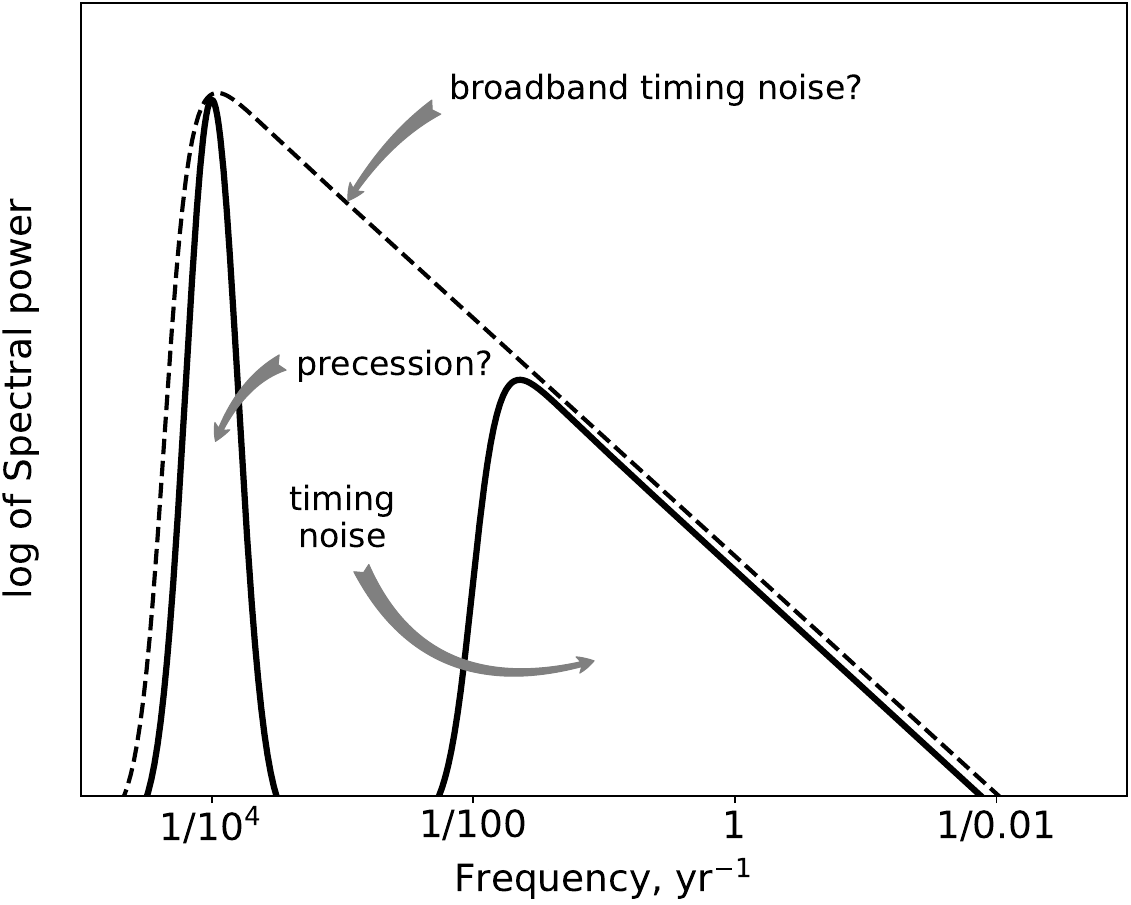}
    \caption{Illustration of the two possible variants of the pulsar timing variations spectrum. The dashed line represents
    a broadband power-law timing noise whose high-frequency part (days-to-months) is directly observed within standard observational time spans.
    We speculate that such a spectrum tends to drop off at time scale $T_\mathrm{max} \sim 10^{4}$ years $< \tau_\mathrm{psr}$, and its low-frequency part can, in principle, dominate the observed braking indices. The solid line represents another option: a spectrum which might consist of two components of a different nature. If this is the case, then only its high-frequency component is related to the common timing noise, while a separate low-frequency component can be connected to the mechanical precession.}
    \label{fig:ssketch}
\end{figure}

The power-law spectrum of the timing noise cannot continue down to zero frequency, and is likely to have a cutoff at 
some timescale $T_\mathrm{max} < \tau_\mathrm{psr}$, as illustrated by the dashed line in Fig.~\ref{fig:ssketch}. In this case, a cyclic component with a characteristic period $T_\mathrm{max}$ will stand out in the long-term spin-down of a pulsar, making its braking index spread within the range of
\begin{equation}
    \Delta n_\mathrm{obs} \sim \frac{\tau_\mathrm{ch}}{T_\mathrm{max}}
\end{equation}
owing to the definition of $n_\mathrm{obs}$. Assuming ergodicity of this process, one gets $T_\mathrm{max} \sim 10^4$~years from the braking indices statistics in Fig.~\ref{fig:ntau}.

On the other hand, the same data show a nearly linear relationship between $\Delta n_\mathrm{obs}$ and $\tau_\mathrm{ch}$ for old pulsars. This could indicate that $T_\mathrm{max}$ is nearly the same for the entire population, in contrast to the variety of timing noise spectra. Moreover, the gray dots in Fig.~\ref{fig:ntau} show that pulsars with {\it intense} timing noise, but with still significant $\ddot\Omega$, follow the same behavior. Therefore, $T_\mathrm{max}$ seems to be independent of the strength of its high-frequency component. This is consistent with the existence of two independent components: a high-frequency one, associated with the common timing noise, and a low-frequency one, probably related to the mechanical precession, accompanied by the magnetic angle oscillations (see the solid line in the Fig.~\ref{fig:ssketch}).

The proposed connection between the braking indices statistics of radiopulsars and their magnetic angle variations allows for direct observational tests. Our model predicts that due to $f(\chi)$ dependence on its argument (see Eq. \ref{eq:f(chi)}), both nearly aligned and nearly orthogonal pulsars must show relatively small positive braking indices. Their deviation from the canonical range $3..3.25$ will be determined mainly by the magnetic field decay. Some indications for the correlation between the magnetic field decay rate $\dot\mu/\mu$ and observed $n_\mathrm{obs}$ have already been discussed by \citet{ip20}. 

Another prediction is that high braking indices should be accompanied by rapid changes in magnetic angle:
\begin{equation}
    \dot\chi \approx -\frac{1}{4f(\chi)} \frac{n_\mathrm{obs}}{\tau_\mathrm{ch}} \approx
    -0.4^\circ\, \left( \frac{n_\mathrm{obs}}{10^5} \right)\left(\frac{\tau_\mathrm{ch}}{10\,\mbox{Myr}}\right)\,\mbox{ yr}^{-1}.
\end{equation}
This rate of the angle evolution is much faster than that predicted by the standard pulsar spin-down model. In principle, it has a 
chance of being detected with polarization observations performed over a time interval of decades.

\subsection{The effect of pinned superfluid}
\label{sec:discuss:super}

Thus far, we treated the NS as a deformed solid body.  In reality, the NS interior, where the density approaches the nuclear density, likely 
contains a neutron superfluid component threaded by an array of quantized vortices \citep{baym69}.  The bulk of the neutron superfluid resides in the core, 
at densities $\rho \gtrsim 2\times10^{14}$ g cm$^{-3}$,  while a small fraction coexists with the crystalline lattice in the inner solid crust.  

The fluid core contains a mixture of superfluid neutrons, superconducting protons and normal electrons. The core plasma is coupled 
to the superfluid through electron scattering off superfluid vortices (which are strongly magnetized owing to entrained proton currents, \citealt{Alpar1984}),
and to the crust through viscous and electromagnetic forces \citep{Alpar87,Alpar1988,jones_anderssson_2001}.  For tight core-crust coupling these two components 
can be considered effectively as a single body \citep{jones_anderssson_2001}.  In practice, this coupling is relatively weak \citep{Alpar1988},
so that free precession is expected to be damped over $\sim 10^3-10^4$ precession periods \citep{Alpar87,jones_anderssson_2001}.  For the radiative precession Eq. \eqref{eq:tau_rad}
implies a damping time longer than the spin down time.  Even for deformations as large as $\varepsilon_{\rm d}\sim 10^{-10}$ the damping time may not be much smaller
then the spin down time, and is typically much longer than the radiative precession time.  Thus, the radiative torque might balance the 
internal torque exerted on the crust by the core, allowing the core to follow the precession of crust \citep{Alpar87}.  Detailed calculations require 
inclusion of the core-crust coupling in a self-consistent manner.

The crust and the superfluid within the crust are thought to be strongly coupled through vortex creep - the pinning and unpinning of vortices to the 
nuclei in the crustal lattice \citep{anderson1975,dalessandro1996,Haskell15,antonelli2022}.   This vortex creep model has become the leading explanation for the `glitch' activity observed in the Crab and Vela pulsars (see, e.g., \citet{Haskell15} and references therein).

As originally shown by \citet{Shaham77} (see also \citealt{jones_anderssson_2001,link2002}), in case of absolute pinning, superfluid neutron vortices, constrained 
to move rigidly with the crust, exert
a large torque on the crust, resulting in rapid precession.   The internal torque exerted on the crust by the pinned superfluid can be added to the 
Euler equations by noting that the angular momentum of the superfluid, henceforth denoted by $\pmb{L}_{s}$, 
is fixed in a frame rotating with the star, viz., $d\pmb{L}_s/dt=0$. With this constraint imposed, Euler equations reduce to
\begin{equation}
	\frac{d\pmb{L}_c}{dt} + \pmb{\Omega} \times ( \pmb{L}_c + \pmb{L}_{s}) = \pmb{N},
    \label{eq:euler}
\end{equation}
where $\pmb{\Omega}$ and $\pmb{L}_c$ are the angular velocity and angular momentum of the crust (and other components tightly coupled to the crust), given by Eqs.~\eqref{eq:Omega} and \eqref{eq:angularL}, respectively, and 
$\pmb{N}$ denotes the external torque acting on the crust (Eq. \ref{eq:torque_decomposed}).   Thus, the dynamics of precession in this case has an additional 
effective torque $\pmb{N}_{s} = -\pmb{\Omega}\times \pmb{L}_{s} = - I_s \, \pmb{\Omega}\times \pmb{\Omega}_s$, where $I_s$ is the moment
of inertia of the pinned superfluid and $\pmb{\Omega}_s$ its angular velocity vector. 
Using  Eq. \eqref{eq:angularL}, we obtain
\begin{equation}
    \pmb{\dot\Omega} + \pmb{\dot\omega} = (\pmb \omega \times \pmb \Omega) + \pmb n 
    - \frac{I_s}{I_0} \, (\pmb{\Omega}\times \pmb{\Omega}_s).
\end{equation}
For values $I_s/I_0 \gtrsim 10^{-2}$ inferred from analysis of glitches and other considerations \citep{link1999,ruderman1976}, we find that the RHS of the latter equation
is largely dominated by the internal torque,
producing a precessional angular frequency $\Omega_p \sim (I_s/I_0)\Omega_s \gtrsim 10^{-2}\Omega$, as originally found by \citet{Shaham77}.

The fast precessional period (minutes to hours)  imposed by absolute pinning is inconsistent with the much longer precessional periods inferred in some neutron stars,
notably Hercules X-1.   It has long been suspected that the superorbital period of 35 days observed in Hercules X-1 is due to (nearly free) precession of the neutron star.
Recent X-ray polarization measurements with the Imaging X-ray Polarimetry Explorer (IXPE), which probe the spin geometry of the neutron star, strongly support this view \citep{Heyl24}.   Nearly free precession might be possible if the internal torque exerted on the crust by the pinned superfluid is balanced by external torques.  In this case a steady state exists in which the pinned superfluid follows the precessional motion of the crust through thermal vortex creep \citep{Alpar87}.
The relaxation time to this equilibrium state depends on the ratio between the pinning and thermal energies, and is on the order of a few hundred to 
a few thousand years for Her X-1.  However, to our understanding \citet{Alpar87} tacitly assumed that the component of the force acting on the superfluid in the direction 
$\pmb{\tilde\omega}\times (\pmb{v}_{\rm n}-\pmb{v}_{\rm s})$, where $\pmb{v}_{\rm n}$ and $\pmb{v}_{\rm s}$ are the velocities of the normal fluid and superfluid, respectively, 
and $\pmb{\tilde\omega}$ is superfluid vorticity, vanishes (see Equations 29 and 51 in \citealt{sedarkin1999}).  Why this is justified is unclear to us.  A more 
general derivation of the torque acting on the superfluid is given in \citet{sedarkin1999}, but their analysis of precession modes is perturbative and
limited to a small (linear) lag around a fixed point where all angular velocities are equal and aligned with a stable principal axis.  
For imperfect pinning, they found low-frequency, strongly damped free precession modes and concluded that the result obtained by \citet{Shaham77} remains intact.
However, it is unclear at present how it would affect the long-term evolution when the lag is large and
the radiative torque is properly taken into account.  In particular, when the lag is close to the critical frequency, one anticipates a relatively large 
rate of vortex motion by thermal creep. 
 
A full treatment of the spin evolution of NS should take into account the coupling between the crust, the core and the superfluid within the crust in a self-consistent manner. Such analysis is beyond the scope of this paper and is left for future work.

\section{Summary}
\label{sec:summary}

In this work, we have conducted a systematic investigation of the long-term rotational evolution of a deformed isolated neutron star, paying particular attention to the dynamics of the magnetic angle $\chi$ (the angle between the rotation and magnetic axes).  We have shown that
in contrast to the gradual monotonic alignment ($\chi \to 0$) found in the case of a spherical star, sufficiently large deformation can lead to high-amplitude cyclic variations of $\chi$ over the precessional time $T_\mathrm{p} \approx 2\pi/\Omega_\mathrm{p}$, with an amplitude that evolves on the spin-down timescale $\tau_\mathrm{psr} \approx I_0\Omega/N_\Omega$.  
We identify several evolutionary regimes distinguished
 by the ratio between the ``effective magnetic deformation'' $\varepsilon_\mu$ (Eq. \ref{eq:epsmu_norm}) and the deformation $\varepsilon_\mathrm{d}$ representing the anisotropy of the star's inertia tensor.
 Most importantly, our model reproduces the peculiar distribution of the pulsar braking indices as a natural consequence of the precession-timescale evolution of a population of realistically deformed pulsars. It can also affect the distribution of magnetic angles. The main evolutionary tracks and
observational consequences can be summarized as follows:

\begin{itemize}
    \item  When $|\varepsilon_\mu| \gg \varepsilon_\mathrm{d}$, the evolution is essentially the same as that of a spherical star; $\chi$  declines monotonically up to a complete alignment of the spin and magnetic axes.
    \item If $|\varepsilon_\mu| \ll \varepsilon_\mathrm{d}$, the evolution of the spin axis depends on the type of deformation, as summarized in Fig. \ref{fig:evol_cheme}.  
    In particular, for triaxial deformation, an attractor state exists in which $\chi$ wobbles with a finite amplitude.
    \item In the intermediate case, $|\varepsilon_\mu| \sim \varepsilon_\mathrm{d}$, the rotation axis either reaches a precession pole or ends up in an attractor orbit.  
    In either case, the spin and magnetic axes remain well separated at all times. 
    %
    %
    \item In situations where the magnetic field decays over a timescale comparable or shorter than the spin-down time, the NS may switch its evolutionary track from a radiative-precession-dominated to one dominated by free precession. In this case, shown in Figure~\ref{fig:phase_triaxial_decay}, the angular size of the precession orbit remains approximately constant. 
    \item At deformation levels $\varepsilon_\mathrm{d} \lesssim 10^{-10}$, the expected oscillations of $\chi$  can account for the abnormally huge braking indices of old pulsars. The observed growth of the braking index $n_\mathrm{obs}$ with the characteristic pulsar age, $\tau_\mathrm{ch} = -\Omega/2\dot\Omega$, is well reproduced for a realistic pulsar population.
    \item Synthetic distributions of $\chi$ computed for a population of deformed stars tend to be uniform in the angle or even isotropic (for the case of triaxial deformations and magnetic field decay).  Compared to the expectations for spherical pulsar populations, they show a lack of nearly aligned and an excess of nearly orthogonal rotators, in good agreement with the observations. 
\end{itemize}

The method developed in this paper for analyzing rotational evolution introduces a new powerful approach to studying and understanding the dynamics and statistics of the magnetic obliquities of isolated neutron stars. However, it treats the star as a deformed solid body and neglects the effect of the superfluid on the evolution, which is expected to 
alter some of the results.  In a follow-up paper we intend to generalize the model to include the coupling of the crust to the crustal superfluid and to the core.

\begin{acknowledgments}
AL thank the Canadian Institute for Theoretical Astrophysics for their warm hospitality and support, and Chris Thompson for vivid discussions.
This work was supported by a grant from the Simons Foundation (00001470), and by the Israel Science Foundation grant 1995/21.
This research was supported in part by grant NSF PHY-2309135 to the Kavli Institute for Theoretical Physics (KITP). 
\end{acknowledgments}

\begin{contribution}

AB and PA performed most of the analytical and numerical analysis and prepared the bulk of the paper. AL provided analysis of the superfluid interiors of NS and wrote Section 5.2. All authors contributed equally to the discussion of intermediate and final results.


\end{contribution}

%

\software{NumPy \citep{harris2020array}, SciPy \citep{2020SciPy-NMeth}, Matplotlib \citep{matplotlib}, Veusz \citep{veusz}
          }


\newpage

\appendix

\section{The evolution equation for the scalar $\lambda$}\label{app:lambda}

Let $\Omega_k = \pmb{\Omega}\cdot \pmb{e}_k$ be the projections of the angular velocity onto the principal axes ($k = 1..3$). 
Then, from the definition of the scalar parameter $\lambda$ (Eq. \ref{eq:lambda_define}), we obtain
\begin{equation}
    2\Omega^2 \lambda = \pmb \Omega \cdot \pmb \Omega_\mathrm{p} = \varepsilon_1 \Omega_1^2 + \varepsilon_3 \Omega_3^2 + \varepsilon_\mu (\pmb \Omega \cdot \pmb m)^2,
\end{equation}
and 
\begin{equation}
    2\frac{\diff}{\diff t}\left(\Omega^2\lambda \right) = (\pmb{\dot\Omega} \cdot \pmb\Omega_\mathrm{p}) + (\pmb\Omega \cdot \pmb{\dot\Omega_\mathrm{p}}).
    \label{eq:dot_lambda_first_step}
\end{equation}
On the other hand, from Eqs. (\ref{eq:omega_radiative}) and (\ref{eq:Omega_precession}), one can write out the components of the total precessional angular velocity in the principal frame as 
\begin{equation}
    \pmb{\Omega}_\mathrm{p} = \left ( \begin{array}{c}
        \varepsilon_1\Omega_1 + \varepsilon_\mu (\pmb\Omega \cdot \pmb m) m_1\\
        \varepsilon_\mu (\pmb\Omega \cdot \pmb m) m_2\\
        \varepsilon_3\Omega_3 + \varepsilon_\mu (\pmb\Omega \cdot \pmb m) m_3\\
    \end{array} \right ),
\end{equation}
where $m_k$ are projections of the unit vector $\pmb m$ onto principal axes. Assuming $\varepsilon_{1,3} = const$, $\varepsilon_\mu = const$, and $m_k = const$, one gets
\begin{equation}
    \pmb \Omega \cdot \pmb{\dot \Omega_\mathrm{p}} = \varepsilon_1 \Omega_1 \dot\Omega_1 + \varepsilon_3 \Omega_3 \dot\Omega_3 + \sum_k \varepsilon_\mu (\pmb{\dot \Omega} \cdot \pmb m) \Omega_k m_k = \frac{\diff}{\diff t} \left ( \Omega^2 \lambda \right),
\end{equation}
as $\sum_k \Omega_k m_k \equiv \pmb \Omega \cdot \pmb m$. 
Taking equation (\ref{eq:dot_lambda_first_step}) into account, one gets
\begin{equation}
    \pmb{\dot \Omega} \cdot \pmb\Omega_\mathrm{p} = \pmb\Omega \cdot \pmb{\dot\Omega}_\mathrm{p}.
\end{equation}
This allows to rewrite the equation (\ref{eq:dot_lambda_first_step}) as
\begin{equation}
    2\Omega \dot \Omega \lambda + \Omega^2 \dot \lambda = (\pmb{\dot\Omega} \cdot \pmb\Omega_\mathrm{p}).
    \label{eq:dot_lambda_second_step}
\end{equation}
Let us first project Euler's equations (\ref{eq:Euler_final}) onto the spin axis $\pmb s_3$:
\begin{equation}
    (1 + \varepsilon_1) \dot\Omega_1 \frac{\Omega_1}{\Omega } + \dot\Omega_2 \frac{\Omega_2}{\Omega} + (1 + \varepsilon_3) \dot\Omega_3 \frac{\Omega_3}{\Omega} = n_\Omega,
\end{equation}
which in case of a slightly deformed star ($|\varepsilon_{1,3}| \ll 1$)
gives
\begin{equation}
    (\pmb{\dot \Omega} \cdot \pmb s_3) = \dot\Omega \approx n_\Omega.
    \label{eq:dot_omega_torque}
\end{equation}
On the other hand, if $\varepsilon_k = const$,
\begin{equation}
    \pmb{\dot\Omega} \cdot \pmb\Omega_\mathrm{p} + \pmb{\dot\omega}\cdot \pmb\Omega_p = (1 + \varepsilon_1) \dot \Omega_1 \Omega_\mathrm{p,1} + \dot\Omega_2 \Omega_\mathrm{p,2} + (1 + \varepsilon_3)\dot\Omega_3 \Omega_\mathrm{p,3} \approx \pmb{\dot\Omega} \cdot \pmb{\Omega}_\mathrm{p}.
    \label{eq:no_small_omega}
\end{equation}
Thus, the Euler equations projected onto the precessional axis $\pmb \Omega_\mathrm{p}$ may be rewritten as
\begin{equation}
    2\Omega\dot\Omega \lambda + \Omega^2\dot\lambda = n_\Omega (\pmb\Omega_\mathrm{p} \cdot \pmb s_3) + n_\chi (\pmb\Omega_\mathrm{p} \cdot \pmb s_2),
\end{equation}
where
\begin{equation}
    n_\Omega(\pmb \Omega_\mathrm{p} \cdot \pmb s_3) = \dot\Omega \left ( \varepsilon_1 \frac{\Omega_1^2}{\Omega} + \varepsilon_3 \frac{\Omega_3^2}{\Omega} + \varepsilon_\mu \frac{(\pmb \Omega \cdot \pmb m)^2}{\Omega}\right) = 2\dot\Omega\Omega\lambda.
\end{equation}
Finally, one obtains the evolutionary equation for $\lambda$ under the assumption of constant deformations $\varepsilon_k$ and fixed $\pmb m$ in a form
\begin{equation}
    \Omega^2 \dot\lambda = n_\chi (\pmb{\Omega}_\mathrm{p} \cdot \pmb s_2) = \pmb\Omega_\mathrm{p} \cdot \pmb n_\chi.
\end{equation}
Then, recapping that $n_\chi = n_\mathrm{psr}\sin\chi\cos\chi$ (Eq.~\ref{eq:N_chi_iso}) and $\pmb s_2\sin\chi = {\pmb s}_3\cos\chi - \pmb m$, (Eq.~\ref{eq:s2_define}) one gets
\begin{equation}
    \Omega^2 \dot\lambda = n_\mathrm{psr} \Omega\lambda \cos^2\chi - n_\mathrm{psr} (\pmb \Omega_\mathrm{p} \cdot \pmb m)\cos\chi,
\end{equation}
or
\begin{equation}
    \tau_\mathrm{psr}\dot\lambda = \varepsilon_\mu \sin^2\chi\cos^2\chi + \varepsilon_1 \frac{\Omega_1}{\Omega}\left (m_1\cos\chi - \frac{\Omega_1}{\Omega}\cos^2\chi\right ) + \varepsilon_3\frac{\Omega_3}{\Omega} \left (m_3\cos\chi - \frac{\Omega_3}{\Omega}\cos^2\chi\right ),
\end{equation}
where $\tau_\mathrm{psr} = -\Omega/n_\mathrm{psr}$ is the spin-down timescale. In another form, if the definition of $\lambda$ is taken into account:
\begin{equation}
    \tau_\mathrm{psr} \frac{\diff \lambda}{\diff t} = (\varepsilon_\mu - 2\lambda)\cos^2\chi + \left(\varepsilon_1 \frac{\Omega_1}{\Omega} m_1 + \varepsilon_3 \frac{\Omega_3}{\Omega} m_3 \right)\cos\chi,
\end{equation}
This equation is useful for analytically considering the precession of an isolated neutron star.


\bibliography{eman}{}

\begin{thebibliography}{}
\expandafter\ifx\csname natexlab\endcsname\relax\def\natexlab#1{#1}\fi
\providecommand{\url}[1]{\href{#1}{#1}}
\providecommand{\dodoi}[1]{doi:~\href{http://doi.org/#1}{\nolinkurl{#1}}}
\providecommand{\doeprint}[1]{\href{http://ascl.net/#1}{\nolinkurl{http://ascl.net/#1}}}
\providecommand{\doarXiv}[1]{\href{https://arxiv.org/abs/#1}{\nolinkurl{https://arxiv.org/abs/#1}}}

\bibitem[{P. {Abolmasov} {et~al.}(2024){Abolmasov}, {Biryukov}, \& {Popov}}]{2024Galax..12....7A}
{Abolmasov}, P., {Biryukov}, A., \& {Popov}, S.~B. 2024, \bibinfo{title}{{Spin Evolution of Neutron Stars},} Galaxies, 12, 7, \dodoi{10.3390/galaxies12010007}

\bibitem[{D.~N. {Aguilera} {et~al.}(2008){Aguilera}, {Pons}, \& {Miralles}}]{agu08}
{Aguilera}, D.~N., {Pons}, J.~A., \& {Miralles}, J.~A. 2008, \bibinfo{title}{{2D Cooling of magnetized neutron stars},} Astron. and Astrophys., 486, 255, \dodoi{10.1051/0004-6361:20078786}

\bibitem[{T. {Akg{\"u}n} {et~al.}(2006){Akg{\"u}n}, {Link}, \& {Wasserman}}]{akgun06}
{Akg{\"u}n}, T., {Link}, B., \& {Wasserman}, I. 2006, \bibinfo{title}{{Precession of the isolated neutron star PSR B1828-11},} \mnras, 365, 653, \dodoi{10.1111/j.1365-2966.2005.09745.x}

\bibitem[{A. {Alpar} \& H. {Oegelman}(1987){Alpar} \& {Oegelman}}]{Alpar87}
{Alpar}, A., \& {Oegelman}, H. 1987, \bibinfo{title}{{Neutron star procession and the dynamics of the superfluid interior.},} Astron. Astrophys., 185, 196

\bibitem[{M.~A. {Alpar} {et~al.}(1984){Alpar}, {Langer}, \& {Sauls}}]{Alpar1984}
{Alpar}, M.~A., {Langer}, S.~A., \& {Sauls}, J.~A. 1984, \bibinfo{title}{{Rapid postglitch spin-up of the superfluid core in pulsars.},} \apj, 282, 533, \dodoi{10.1086/162232}

\bibitem[{M.~A. {Alpar} \& J.~A. {Sauls}(1988){Alpar} \& {Sauls}}]{Alpar1988}
{Alpar}, M.~A., \& {Sauls}, J.~A. 1988, \bibinfo{title}{{On the Dynamical Coupling between the Superfluid Interior and the Crust of a Neutron Star},} \apj, 327, 723, \dodoi{10.1086/166228}

\bibitem[{P.~W. {Anderson} \& N. {Itoh}(1975){Anderson} \& {Itoh}}]{anderson1975}
{Anderson}, P.~W., \& {Itoh}, N. 1975, \bibinfo{title}{{Pulsar glitches and restlessness as a hard superfluidity phenomenon},} Nature, 256, 25, \dodoi{10.1038/256025a0}

\bibitem[{M. {Antonelli} {et~al.}(2022){Antonelli}, {Montoli}, \& {Pizzochero}}]{antonelli2022}
{Antonelli}, M., {Montoli}, A., \& {Pizzochero}, P.~M. 2022, \bibinfo{title}{{Insights Into the Physics of Neutron Star Interiors from Pulsar Glitches},} in Astrophysics in the XXI Century with Compact Stars. Edited by C.A.Z. Vasconcellos. eISBN 978-981-12-2094-4. Singapore: World Scientific, ed. C.~A.~Z. {Vasconcellos}, 219--281, \dodoi{10.1142/9789811220944_0007}

\bibitem[{L. {Arzamasskiy} {et~al.}(2015){Arzamasskiy}, {Philippov}, \& {Tchekhovskoy}}]{arz15}
{Arzamasskiy}, L., {Philippov}, A., \& {Tchekhovskoy}, A. 2015, \bibinfo{title}{{Evolution of non-spherical pulsars with plasma-filled magnetospheres},} MNRAS, 453, 3540, \dodoi{10.1093/mnras/stv1818}

\bibitem[{A. {Baykal} {et~al.}(1999){Baykal}, {Ali Alpar}, {Boynton}, \& {Deeter}}]{1999MNRAS.306..207B}
{Baykal}, A., {Ali Alpar}, M., {Boynton}, P.~E., \& {Deeter}, J.~E. 1999, \bibinfo{title}{{The timing noise of PSR 0823+26,1706-16, 1749-28, 2021+51 and the anomalous brakingindices},} \mnras, 306, 207, \dodoi{10.1046/j.1365-8711.1999.02505.x}

\bibitem[{G. {Baym} {et~al.}(1969){Baym}, {Pethick}, \& {Pines}}]{baym69}
{Baym}, G., {Pethick}, C., \& {Pines}, D. 1969, \bibinfo{title}{{Superfluidity in Neutron Stars},} Nature, 224, 673, \dodoi{10.1038/224673a0}

\bibitem[{V.~S. {Beskin} {et~al.}(1993){Beskin}, {Gurevich}, \& {Istomin}}]{bes93}
{Beskin}, V.~S., {Gurevich}, A.~V., \& {Istomin}, Y.~N. 1993, {Physics of the pulsar magnetosphere} ({Cambridge, New York: Cambridge University Press})

\bibitem[{V.~S. {Beskin} \& A.~A. {Zheltoukhov}(2014){Beskin} \& {Zheltoukhov}}]{bz_anomal2014}
{Beskin}, V.~S., \& {Zheltoukhov}, A.~A. 2014, \bibinfo{title}{{Anomalous torque applied to a rotating magnetized sphere in a vacuum},} Physics Uspekhi, 57, 799, \dodoi{10.3367/UFNe.0184.201408e.0865}

\bibitem[{L. {Bildsten}(1998){Bildsten}}]{Bildsten98}
{Bildsten}, L. 1998, \bibinfo{title}{{Gravitational Radiation and Rotation of Accreting Neutron Stars},} \apjl, 501, L89, \dodoi{10.1086/311440}

\bibitem[{A. {Biryukov} \& G. {Beskin}(2023){Biryukov} \& {Beskin}}]{bb23}
{Biryukov}, A., \& {Beskin}, G. 2023, \bibinfo{title}{{Imprint of magnetic obliquity in apparent spin-down of radio pulsars},} MNRAS, 522, 6258, \dodoi{10.1093/mnras/stad1437}

\bibitem[{A. {Biryukov} {et~al.}(2012){Biryukov}, {Beskin}, \& {Karpov}}]{bbk12}
{Biryukov}, A., {Beskin}, G., \& {Karpov}, S. 2012, \bibinfo{title}{{Monotonic and cyclic components of radio pulsar spin-down},} MNRAS, 420, 103, \dodoi{10.1111/j.1365-2966.2011.20005.x}

\bibitem[{A. {Biryukov} {et~al.}(2007){Biryukov}, {Beskin}, {Karpov}, \& {Chmyreva}}]{bbk07}
{Biryukov}, A., {Beskin}, G., {Karpov}, S., \& {Chmyreva}, L. 2007, \bibinfo{title}{{Evidence of long-term cyclic evolution of radio pulsar periods},} Advances in Space Research, 40, 1498, \dodoi{10.1016/j.asr.2007.06.051}

\bibitem[{P.~E. {Boynton} {et~al.}(1972){Boynton}, {Groth}, {Hutchinson}, {Nanos}, {Partridge}, \& {Wilkinson}}]{boyn72}
{Boynton}, P.~E., {Groth}, E.~J., {Hutchinson}, D.~P., {et~al.} 1972, \bibinfo{title}{{Optical Timing of the Crab Pulsar, NP 0532},} ApJ, 175, 217, \dodoi{10.1086/151550}

\bibitem[{A. {Bransgrove} {et~al.}(2025){Bransgrove}, {Levin}, \& {Beloborodov}}]{bransgrove2025}
{Bransgrove}, A., {Levin}, Y., \& {Beloborodov}, A.~M. 2025, \bibinfo{title}{{Giant Hall Waves Launched by Superconducting Phase Transition in Pulsars},} \apj, 979, 144, \dodoi{10.3847/1538-4357/ad90a3}

\bibitem[{K. {Chen} {et~al.}(1998){Chen}, {Ruderman}, \& {Zhu}}]{chen98}
{Chen}, K., {Ruderman}, M., \& {Zhu}, T. 1998, \bibinfo{title}{{Millisecond Pulsar Alignment: PSR J0437-4715},} Astrophys. J, 493, 397, \dodoi{10.1086/305106}

\bibitem[{A. {Cheng} {et~al.}(1976){Cheng}, {Ruderman}, \& {Sutherland}}]{ruderman1976}
{Cheng}, A., {Ruderman}, M., \& {Sutherland}, P. 1976, \bibinfo{title}{{Current flow in pulsar magnetospheres.},} Astrophys. J, 203, 209, \dodoi{10.1086/154068}

\bibitem[{J.~M. {Cordes} \& D.~J. {Helfand}(1980){Cordes} \& {Helfand}}]{ch80}
{Cordes}, J.~M., \& {Helfand}, D.~J. 1980, \bibinfo{title}{{Pulsar timing. III - Timing noise of 50 pulsars},} \apj, 239, 640, \dodoi{10.1086/158150}

\bibitem[{F. {D'Alessandro}(1996){D'Alessandro}}]{dalessandro1996}
{D'Alessandro}, F. 1996, \bibinfo{title}{{Rotational irregularities in pulsars {\textemdash} A review},} \apss, 246, 73, \dodoi{10.1007/BF00637401}

\bibitem[{F. {D'Alessandro} {et~al.}(1997){D'Alessandro}, {Deshpande}, \& {McCulloch}}]{1997JApA...18....5D}
{D'Alessandro}, F., {Deshpande}, A.~A., \& {McCulloch}, P.~M. 1997, \bibinfo{title}{{Power Spectrum Analysis of the Timing Noise in 18 Southern Pulsars},} Journal of Astrophysics and Astronomy, 18, 5, \dodoi{10.1007/BF02714849}

\bibitem[{F. {D'Alessandro} {et~al.}(1995){D'Alessandro}, {McCulloch}, {Hamilton}, \& {Deshpande}}]{daless95}
{D'Alessandro}, F., {McCulloch}, P.~M., {Hamilton}, P.~A., \& {Deshpande}, A.~A. 1995, \bibinfo{title}{{The timing noise of 45 southern pulsars},} \mnras, 277, 1033, \dodoi{10.1093/mnras/277.3.1033}

\bibitem[{L. {Davis} \& M. {Goldstein}(1970){Davis} \& {Goldstein}}]{dg70}
{Davis}, L., \& {Goldstein}, M. 1970, \bibinfo{title}{{Magnetic-Dipole Alignment in Pulsars},} Astrophys. J. Lett., 159, \dodoi{10.1086/180482}

\bibitem[{P. {Goldreich}(1970){Goldreich}}]{goldreich70}
{Goldreich}, P. 1970, \bibinfo{title}{{Neutron Star Crusts and Alignment of Magnetic Axes in Pulsars},} \apjl, 160, L11, \dodoi{10.1086/180513}

\bibitem[{M.~L. {Good} \& K.~K. {Ng}(1985){Good} \& {Ng}}]{good85}
{Good}, M.~L., \& {Ng}, K.~K. 1985, \bibinfo{title}{{Electromagnetic torques, secular alignment, and spin-down of neutron stars},} Astrophys. J., 299, 706, \dodoi{10.1086/163736}

\bibitem[{K.~N. {Gourgouliatos} \& A. {Cumming}(2014{\natexlab{a}}){Gourgouliatos} \& {Cumming}}]{gc14prl}
{Gourgouliatos}, K.~N., \& {Cumming}, A. 2014{\natexlab{a}}, \bibinfo{title}{{Hall Attractor in Axially Symmetric Magnetic Fields in Neutron Star Crusts},} \prl, 112, 171101, \dodoi{10.1103/PhysRevLett.112.171101}

\bibitem[{K.~N. {Gourgouliatos} \& A. {Cumming}(2014{\natexlab{b}}){Gourgouliatos} \& {Cumming}}]{gc14mnras}
{Gourgouliatos}, K.~N., \& {Cumming}, A. 2014{\natexlab{b}}, \bibinfo{title}{{Hall effect in neutron star crusts: evolution, endpoint and dependence on initial conditions},} \mnras, 438, 1618, \dodoi{10.1093/mnras/stt2300}

\bibitem[{C.~R. Harris {et~al.}(2020)Harris, Millman, van~der Walt, Gommers, Virtanen, Cournapeau, Wieser, Taylor, Berg, Smith, Kern, Picus, Hoyer, van Kerkwijk, Brett, Haldane, del R{\'{i}}o, Wiebe, Peterson, G{\'{e}}rard-Marchant, Sheppard, Reddy, Weckesser, Abbasi, Gohlke, \& Oliphant}]{harris2020array}
Harris, C.~R., Millman, K.~J., van~der Walt, S.~J., {et~al.} 2020, \bibinfo{title}{Array programming with {NumPy},} Nature, 585, 357, \dodoi{10.1038/s41586-020-2649-2}

\bibitem[{B. {Haskell} \& A. {Melatos}(2015){Haskell} \& {Melatos}}]{Haskell15}
{Haskell}, B., \& {Melatos}, A. 2015, \bibinfo{title}{{Models of pulsar glitches},} International Journal of Modern Physics D, 24, 1530008, \dodoi{10.1142/S0218271815300086}

\bibitem[{B. {Haskell} {et~al.}(2008){Haskell}, {Samuelsson}, {Glampedakis}, \& {Andersson}}]{haskell08}
{Haskell}, B., {Samuelsson}, L., {Glampedakis}, K., \& {Andersson}, N. 2008, \bibinfo{title}{{Modelling magnetically deformed neutron stars},} MNRAS, 385, 531, \dodoi{10.1111/j.1365-2966.2008.12861.x}

\bibitem[{J. {Heyl} {et~al.}(2024){Heyl}, {Doroshenko}, {Gonz{\'a}lez-Caniulef}, {Caiazzo}, {Poutanen}, {Mushtukov}, {Tsygankov}, {Kirmizibayrak}, {Bachetti}, {Pavlov}, {Forsblom}, {Malacaria}, {Suleimanov}, {Agudo}, {Antonelli}, {Baldini}, {Baumgartner}, {Bellazzini}, {Bianchi}, {Bongiorno}, {Bonino}, {Brez}, {Bucciantini}, {Capitanio}, {Castellano}, {Cavazzuti}, {Chen}, {Ciprini}, {Costa}, {De Rosa}, {Del Monte}, {Di Gesu}, {Di Lalla}, {Di Marco}, {Donnarumma}, {Dov{\v{c}}iak}, {Ehlert}, {Enoto}, {Evangelista}, {Fabiani}, {Ferrazzoli}, {Garcia}, {Gunji}, {Hayashida}, {Iwakiri}, {Jorstad}, {Kaaret}, {Karas}, {Kislat}, {Kitaguchi}, {Kolodziejczak}, {Krawczynski}, {La Monaca}, {Latronico}, {Liodakis}, {Maldera}, {Manfreda}, {Marin}, {Marinucci}, {Marscher}, {Marshall}, {Massaro}, {Matt}, {Mitsuishi}, {Mizuno}, {Muleri}, {Negro}, {Ng}, {O'Dell}, {Omodei}, {Oppedisano}, {Papitto}, {Peirson}, {Perri}, {Pesce-Rollins}, {Petrucci}, {Pilia}, {Possenti}, {Puccetti}, {Ramsey}, {Rankin}, {Ratheesh}, {Roberts},
  {Romani}, {Sgr{\`o}}, {Slane}, {Soffitta}, {Spandre}, {Swartz}, {Tamagawa}, {Tavecchio}, {Taverna}, {Tawara}, {Tennant}, {Thomas}, {Tombesi}, {Trois}, {Turolla}, {Vink}, {Weisskopf}, {Wu}, {Xie}, \& {Zane}}]{Heyl24}
{Heyl}, J., {Doroshenko}, V., {Gonz{\'a}lez-Caniulef}, D., {et~al.} 2024, \bibinfo{title}{{Complex rotational dynamics of the neutron star in Hercules X-1 revealed by X-ray polarization},} Nature Astronomy, 8, 1047, \dodoi{10.1038/s41550-024-02295-8}

\bibitem[{G. {Hobbs} {et~al.}(2010){Hobbs}, {Lyne}, \& {Kramer}}]{hobbs10}
{Hobbs}, G., {Lyne}, A.~G., \& {Kramer}, M. 2010, \bibinfo{title}{{An analysis of the timing irregularities for 366 pulsars},} MNRAS, 402, 1027, \dodoi{10.1111/j.1365-2966.2009.15938.x}

\bibitem[{J.~D. Hunter(2007)Hunter}]{matplotlib}
Hunter, J.~D. 2007, \bibinfo{title}{Matplotlib: A 2D graphics environment,} Computing in Science \& Engineering, 9, 90, \dodoi{10.1109/MCSE.2007.55}

\bibitem[{A.~P. {Igoshev}(2019){Igoshev}}]{Igoshev19}
{Igoshev}, A.~P. 2019, \bibinfo{title}{{Ages of radio pulsar: long-term magnetic field evolution},} \mnras, 482, 3415, \dodoi{10.1093/mnras/sty2945}

\bibitem[{A.~P. {Igoshev} {et~al.}(2022){Igoshev}, {Frantsuzova}, {Gourgouliatos}, {Tsichli}, {Konstantinou}, \& {Popov}}]{Igoshev22}
{Igoshev}, A.~P., {Frantsuzova}, A., {Gourgouliatos}, K.~N., {et~al.} 2022, \bibinfo{title}{{Initial periods and magnetic fields of neutron stars},} MNRAS, 514, 4606, \dodoi{10.1093/mnras/stac1648}

\bibitem[{A.~P. {Igoshev} \& S.~B. {Popov}(2020){Igoshev} \& {Popov}}]{ip20}
{Igoshev}, A.~P., \& {Popov}, S.~B. 2020, \bibinfo{title}{{Braking indices of young radio pulsars: theoretical perspective},} \mnras, 499, 2826, \dodoi{10.1093/mnras/staa3070}

\bibitem[{A.~P. {Igoshev} {et~al.}(2021){Igoshev}, {Popov}, \& {Hollerbach}}]{ip21}
{Igoshev}, A.~P., {Popov}, S.~B., \& {Hollerbach}, R. 2021, \bibinfo{title}{{Evolution of Neutron Star Magnetic Fields},} Universe, 7, 351, \dodoi{10.3390/universe7090351}

\bibitem[{T.~J. {Johnson} {et~al.}(2014){Johnson}, {Venter}, {Harding}, {Guillemot}, {Smith}, {Kramer}, {{\c{C}}elik}, {den Hartog}, {Ferrara}, \& {Hou}}]{johnson14}
{Johnson}, T.~J., {Venter}, C., {Harding}, A.~K., {et~al.} 2014, \bibinfo{title}{{Constraints on the Emission Geometries and Spin Evolution of Gamma-Ray Millisecond Pulsars},} Astrophys. J. Suppl., 213, 6, \dodoi{10.1088/0067-0049/213/1/6}

\bibitem[{D.~I. {Jones} \& N. {Andersson}(2001){Jones} \& {Andersson}}]{jones_anderssson_2001}
{Jones}, D.~I., \& {Andersson}, N. 2001, \bibinfo{title}{{Freely precessing neutron stars: model and observations},} \mnras, 324, 811, \dodoi{10.1046/j.1365-8711.2001.04251.x}

\bibitem[{F.~A. {Kniazev} {et~al.}(2025){Kniazev}, {Istomin}, \& {Beskin}}]{Kniazev25}
{Kniazev}, F.~A., {Istomin}, A.~Y., \& {Beskin}, V.~S. 2025, \bibinfo{title}{{Data from FAST and MeerKAT surveys as a test of radio pulsar physics},} arXiv e-prints, arXiv:2506.12423, \dodoi{10.48550/arXiv.2506.12423}

\bibitem[{D. {Kolesnikov} {et~al.}(2022){Kolesnikov}, {Shakura}, \& {Postnov}}]{Kolesnikov22}
{Kolesnikov}, D., {Shakura}, N., \& {Postnov}, K. 2022, \bibinfo{title}{{Evidence for neutron star triaxial free precession in Her X-1 from Fermi/GBM pulse period measurements},} \mnras, 513, 3359, \dodoi{10.1093/mnras/stac1107}

\bibitem[{Y. {Levin} {et~al.}(2020){Levin}, {Beloborodov}, \& {Bransgrove}}]{Levin20}
{Levin}, Y., {Beloborodov}, A.~M., \& {Bransgrove}, A. 2020, \bibinfo{title}{{Precessing Flaring Magnetar as a Source of Repeating FRB 180916.J0158+65},} \apjl, 895, L30, \dodoi{10.3847/2041-8213/ab8c4c}

\bibitem[{B. {Link} \& C. {Cutler}(2002){Link} \& {Cutler}}]{link2002}
{Link}, B., \& {Cutler}, C. 2002, \bibinfo{title}{{Vortex unpinning in precessing neutron stars},} MNRAS, 336, 211, \dodoi{10.1046/j.1365-8711.2002.05726.x}

\bibitem[{B. {Link} {et~al.}(1999){Link}, {Epstein}, \& {Lattimer}}]{link1999}
{Link}, B., {Epstein}, R.~I., \& {Lattimer}, J.~M. 1999, \bibinfo{title}{{Pulsar Constraints on Neutron Star Structure and Equation of State},} Phys. Rev. Lett, 83, 3362, \dodoi{10.1103/PhysRevLett.83.3362}

\bibitem[{A. {Lyne} \& F. {Graham-Smith}(2012){Lyne} \& {Graham-Smith}}]{2012puas.book.....L}
{Lyne}, A., \& {Graham-Smith}, F. 2012, {Pulsar Astronomy}

\bibitem[{A. {Lyne} {et~al.}(2013){Lyne}, {Graham-Smith}, {Weltevrede}, {Jordan}, {Stappers}, {Bassa}, \& {Kramer}}]{lyne13}
{Lyne}, A., {Graham-Smith}, F., {Weltevrede}, P., {et~al.} 2013, \bibinfo{title}{{Evolution of the Magnetic Field Structure of the Crab Pulsar},} Science, 342, 598, \dodoi{10.1126/science.1243254}

\bibitem[{A.~G. {Lyne} \& R.~N. {Manchester}(1988){Lyne} \& {Manchester}}]{lyne88}
{Lyne}, A.~G., \& {Manchester}, R.~N. 1988, \bibinfo{title}{{The shape of pulsar radio beams},} MNRAS, 234, 477, \dodoi{10.1093/mnras/234.3.477}

\bibitem[{I.~F. {Malov} \& E.~B. {Nikitina}(2011){Malov} \& {Nikitina}}]{nikitina11}
{Malov}, I.~F., \& {Nikitina}, E.~B. 2011, \bibinfo{title}{{The geometry of radio pulsar magnetospheres},} Astronomy Reports, 55, 878, \dodoi{10.1134/S1063772911100076}

\bibitem[{R.~N. {Manchester} {et~al.}(2005){Manchester}, {Hobbs}, {Teoh}, \& {Hobbs}}]{atnf}
{Manchester}, R.~N., {Hobbs}, G.~B., {Teoh}, A., \& {Hobbs}, M. 2005, \bibinfo{title}{{The Australia Telescope National Facility Pulsar Catalogue},} AJ, 129, 1993, \dodoi{10.1086/428488}

\bibitem[{A. {Melatos}(2000){Melatos}}]{melatos2000}
{Melatos}, A. 2000, \bibinfo{title}{{Radiative precession of an isolated neutron star},} MNRAS, 313, 217, \dodoi{10.1046/j.1365-8711.2000.03031.x}

\bibitem[{E.~M. {Novoselov} {et~al.}(2020){Novoselov}, {Beskin}, {Galishnikova}, {Rashkovetskyi}, \& {Biryukov}}]{Novoselov2020}
{Novoselov}, E.~M., {Beskin}, V.~S., {Galishnikova}, A.~K., {Rashkovetskyi}, M.~M., \& {Biryukov}, A.~V. 2020, \bibinfo{title}{{Orthogonal pulsars as a key test for pulsar evolution},} MNRAS, 494, 3899, \dodoi{10.1093/mnras/staa904}

\bibitem[{J.~P. {Ostriker} \& J.~E. {Gunn}(1969){Ostriker} \& {Gunn}}]{ogunn69}
{Ostriker}, J.~P., \& {Gunn}, J.~E. 1969, \bibinfo{title}{{On the Nature of Pulsars. I. Theory},} Astrophys. J., 157, 1395, \dodoi{10.1086/150160}

\bibitem[{Z.~W. {Ou} {et~al.}(2016){Ou}, {Tong}, {Kou}, \& {Ding}}]{ou16}
{Ou}, Z.~W., {Tong}, H., {Kou}, F.~F., \& {Ding}, G.~Q. 2016, \bibinfo{title}{{Fluctuating neutron star magnetosphere: braking indices of eight pulsars, frequency second derivatives of 222 pulsars and 15 magnetars},} \mnras, 457, 3922, \dodoi{10.1093/mnras/stw227}

\bibitem[{J. {P{\'e}tri}(2020){P{\'e}tri}}]{Petri20}
{P{\'e}tri}, J. 2020, \bibinfo{title}{{Electrodynamics and Radiation from Rotating Neutron Star Magnetospheres},} Universe, 6, 15, \dodoi{10.3390/universe6010015}

\bibitem[{A. {Philippov} {et~al.}(2014){Philippov}, {Tchekhovskoy}, \& {Li}}]{phil14}
{Philippov}, A., {Tchekhovskoy}, A., \& {Li}, J.~G. 2014, \bibinfo{title}{{Time evolution of pulsar obliquity angle from 3D simulations of magnetospheres},} MNRAS, 441, 1879, \dodoi{10.1093/mnras/stu591}

\bibitem[{A.~A. {Philippov} {et~al.}(2015){Philippov}, {Spitkovsky}, \& {Cerutti}}]{phil15}
{Philippov}, A.~A., {Spitkovsky}, A., \& {Cerutti}, B. 2015, \bibinfo{title}{{Ab Initio Pulsar Magnetosphere: Three-dimensional Particle-in-cell Simulations of Oblique Pulsars},} Astrophys. J. Lett., 801, L19, \dodoi{10.1088/2041-8205/801/1/L19}

\bibitem[{J.~A. {Pons} {et~al.}(2007){Pons}, {Link}, {Miralles}, \& {Geppert}}]{pons07}
{Pons}, J.~A., {Link}, B., {Miralles}, J.~A., \& {Geppert}, U. 2007, \bibinfo{title}{{Evidence for Heating of Neutron Stars by Magnetic-Field Decay},} Physical Review Letters, 98, 071101, \dodoi{10.1103/PhysRevLett.98.071101}

\bibitem[{J.~A. {Pons} {et~al.}(2009){Pons}, {Miralles}, \& {Geppert}}]{pons09}
{Pons}, J.~A., {Miralles}, J.~A., \& {Geppert}, U. 2009, \bibinfo{title}{{Magneto-thermal evolution of neutron stars},} \aap, 496, 207, \dodoi{10.1051/0004-6361:200811229}

\bibitem[{J.~A. {Pons} {et~al.}(2012){Pons}, {Vigan{\`o}}, \& {Geppert}}]{pons12}
{Pons}, J.~A., {Vigan{\`o}}, D., \& {Geppert}, U. 2012, \bibinfo{title}{{Pulsar timing irregularities and the imprint of magnetic field evolution},} Astron. and Astrophys., 547, A9, \dodoi{10.1051/0004-6361/201220091}

\bibitem[{J.~M. {Rankin}(1990){Rankin}}]{rankin90}
{Rankin}, J.~M. 1990, \bibinfo{title}{{Toward an Empirical Theory of Pulsar Emission. IV. Geometry of the Core Emission Region},} Astrophys. J, 352, 247, \dodoi{10.1086/168530}

\bibitem[{J. {Sanders}(2023){Sanders}}]{veusz}
{Sanders}, J. 2023, {Veusz: Scientific plotting package},, Astrophysics Source Code Library, record ascl:2307.017

\bibitem[{A. {Sedrakian} {et~al.}(1999){Sedrakian}, {Wasserman}, \& {Cordes}}]{sedarkin1999}
{Sedrakian}, A., {Wasserman}, I., \& {Cordes}, J.~M. 1999, \bibinfo{title}{{Precession of Isolated Neutron Stars. I. Effects of Imperfect Pinning},} \apj, 524, 341, \dodoi{10.1086/307777}

\bibitem[{T.~V. {Shabanova} {et~al.}(2001){Shabanova}, {Lyne}, \& {Urama}}]{Shabanova01}
{Shabanova}, T.~V., {Lyne}, A.~G., \& {Urama}, J.~O. 2001, \bibinfo{title}{{Evidence for Free Precession in the Pulsar B1642-03},} \apj, 552, 321, \dodoi{10.1086/320438}

\bibitem[{J. {Shaham}(1977){Shaham}}]{Shaham77}
{Shaham}, J. 1977, \bibinfo{title}{{Free precession of neutron stars: role of possible vortex pinning.},} Astrophys. J, 214, 251, \dodoi{10.1086/155249}

\bibitem[{P. {Soffitta} {et~al.}(2021){Soffitta}, {Baldini}, {Bellazzini}, {Costa}, {Latronico}, {Muleri}, {Del Monte}, {Fabiani}, {Minuti}, {Pinchera}, {Sgro'}, {Spandre}, {Trois}, {Amici}, {Andersson}, {Attina'}, {Bachetti}, {Barbanera}, {Borotto}, {Brez}, {Brienza}, {Caporale}, {Cardelli}, {Carpentiero}, {Castellano}, {Castronuovo}, {Cavalli}, {Cavazzuti}, {Ceccanti}, {Centrone}, {Ciprini}, {Citraro}, {D'Amico}, {D'Alba}, {Di Cosimo}, {Di Lalla}, {Di Marco}, {Di Persio}, {Donnarumma}, {Evangelista}, {Ferrazzoli}, {Hayato}, {Kitaguchi}, {La Monaca}, {Lefevre}, {Loffredo}, {Lorenzi}, {Lucchesi}, {Magazzu}, {Maldera}, {Manfreda}, {Mangraviti}, {Marengo}, {Matt}, {Mereu}, {Morbidini}, {Mosti}, {Nakano}, {Nasimi}, {Negri}, {Nenonen}, {Nuti}, {Orsini}, {Perri}, {Pesce-Rollins}, {Piazzolla}, {Pilia}, {Profeti}, {Puccetti}, {Rankin}, {Ratheesh}, {Rubini}, {Santoli}, {Sarra}, {Scalise}, {Sciortino}, {Tamagawa}, {Tardiola}, {Tobia}, {Vimercati}, \& {Xie}}]{2021AJ....162..208S}
{Soffitta}, P., {Baldini}, L., {Bellazzini}, R., {et~al.} 2021, \bibinfo{title}{{The Instrument of the Imaging X-Ray Polarimetry Explorer},} \aj, 162, 208, \dodoi{10.3847/1538-3881/ac19b0}

\bibitem[{A. {Spitkovsky}(2006){Spitkovsky}}]{spitkovsky06}
{Spitkovsky}, A. 2006, \bibinfo{title}{{Time-dependent Force-free Pulsar Magnetospheres: Axisymmetric and Oblique Rotators},} Astrophys. J. Lett., 648, L51, \dodoi{10.1086/507518}

\bibitem[{I.~H. {Stairs} {et~al.}(2019){Stairs}, {Lyne}, {Kramer}, {Stappers}, {van Leeuwen}, {Tung}, {Manchester}, {Hobbs}, {Lorimer}, \& {Melatos}}]{stairs19}
{Stairs}, I.~H., {Lyne}, A.~G., {Kramer}, M., {et~al.} 2019, \bibinfo{title}{{Mode switching and oscillations in PSR B1828-11},} \mnras, 485, 3230, \dodoi{10.1093/mnras/stz647}

\bibitem[{T.~M. {Tauris} \& R.~N. {Manchester}(1998){Tauris} \& {Manchester}}]{tm98}
{Tauris}, T.~M., \& {Manchester}, R.~N. 1998, \bibinfo{title}{{On the Evolution of Pulsar Beams},} MNRAS, 298, 625, \dodoi{10.1046/j.1365-8711.1998.01369.x}

\bibitem[{A. {Tchekhovskoy} {et~al.}(2013){Tchekhovskoy}, {Spitkovsky}, \& {Li}}]{tche13}
{Tchekhovskoy}, A., {Spitkovsky}, A., \& {Li}, J.~G. 2013, \bibinfo{title}{{Time-dependent 3D magnetohydrodynamic pulsar magnetospheres: oblique rotators},} MNRAS, 435, L1, \dodoi{10.1093/mnrasl/slt076}

\bibitem[{J.~O. {Urama} {et~al.}(2006){Urama}, {Link}, \& {Weisberg}}]{urama06}
{Urama}, J.~O., {Link}, B., \& {Weisberg}, J.~M. 2006, \bibinfo{title}{{A strong correlation in radio pulsars with implications for torque variations},} \mnras, 370, L76, \dodoi{10.1111/j.1745-3933.2006.00192.x}

\bibitem[{G. {Ushomirsky} {et~al.}(2000){Ushomirsky}, {Cutler}, \& {Bildsten}}]{usho2000}
{Ushomirsky}, G., {Cutler}, C., \& {Bildsten}, L. 2000, \bibinfo{title}{{Deformations of accreting neutron star crusts and gravitational wave emission},} \mnras, 319, 902, \dodoi{10.1046/j.1365-8711.2000.03938.x}

\bibitem[{P. Virtanen {et~al.}(2020)Virtanen, Gommers, Oliphant, Haberland, Reddy, Cournapeau, Burovski, Peterson, Weckesser, Bright, {van der Walt}, Brett, Wilson, Millman, Mayorov, Nelson, Jones, Kern, Larson, Carey, Polat, Feng, Moore, {VanderPlas}, Laxalde, Perktold, Cimrman, Henriksen, Quintero, Harris, Archibald, Ribeiro, Pedregosa, {van Mulbregt}, \& {SciPy 1.0 Contributors}}]{2020SciPy-NMeth}
Virtanen, P., Gommers, R., Oliphant, T.~E., {et~al.} 2020, \bibinfo{title}{{{SciPy} 1.0: Fundamental Algorithms for Scientific Computing in Python},} Nature Methods, 17, 261, \dodoi{10.1038/s41592-019-0686-2}

\bibitem[{P.~F. {Wang} {et~al.}(2023){Wang}, {Han}, {Xu}, {Wang}, {Yan}, {Jing}, {Su}, {Zhou}, \& {Wang}}]{FAST_polarization23}
{Wang}, P.~F., {Han}, J.~L., {Xu}, J., {et~al.} 2023, \bibinfo{title}{{FAST Pulsar Database. I. Polarization Profiles of 682 Pulsars},} Research in Astronomy and Astrophysics, 23, 104002, \dodoi{10.1088/1674-4527/acea1f}

\bibitem[{I. {Wasserman} {et~al.}(2022){Wasserman}, {Cordes}, {Chatterjee}, \& {Batra}}]{wasserman2022}
{Wasserman}, I., {Cordes}, J.~M., {Chatterjee}, S., \& {Batra}, G. 2022, \bibinfo{title}{{Nonaxisymmetric Precession of Magnetars and Fast Radio Bursts},} \apj, 928, 53, \dodoi{10.3847/1538-4357/ac38a6}

\bibitem[{J. {Yuan} {et~al.}(2024){Yuan}, {Wang}, {Dang}, {Li}, {Kou}, {Yan}, {Wen}, {Liu}, {Yuen}, {Wang}, {Zhou}, {Liu}, \& {He}}]{2024Univ...10..105Y}
{Yuan}, J., {Wang}, N., {Dang}, S., {et~al.} 2024, \bibinfo{title}{{Characterizing Timing Noise in Normal Pulsars with the Nanshan Radio Telescope},} Universe, 10, 105, \dodoi{10.3390/universe10030105}

\bibitem[{J.~J. {Zanazzi} \& D. {Lai}(2015){Zanazzi} \& {Lai}}]{zanazzi15}
{Zanazzi}, J.~J., \& {Lai}, D. 2015, \bibinfo{title}{{Electromagnetic torques, precession and evolution of magnetic inclination of pulsars},} \mnras, 451, 695, \dodoi{10.1093/mnras/stv955}

\bibitem[{B. {Zhang} {et~al.}(2024){Zhang}, {Zhong}, {Li}, \& {Dai}}]{zhang24}
{Zhang}, B., {Zhong}, S.-Q., {Li}, L., \& {Dai}, Z.-G. 2024, \bibinfo{title}{{Signature of Triaxially Precessing Magnetars in Gamma-Ray Burst X-Ray Afterglows},} \apj, 977, 206, \dodoi{10.3847/1538-4357/ad9005}

\bibitem[{S.-N. {Zhang} \& Y. {Xie}(2012){Zhang} \& {Xie}}]{zhang12}
{Zhang}, S.-N., \& {Xie}, Y. 2012, \bibinfo{title}{{Why Do the Braking Indices of Pulsars Span a Range of More Than 100 Millions?},} Astrophys. J., 761, 102, \dodoi{10.1088/0004-637X/761/2/102}

\end{thebibliography}
\bibliographystyle{aasjournalv7}



\end{document}